\documentclass[11pt]{article}
\usepackage{jheppub}
\usepackage{tensor}
\usepackage{cleveref}
\usepackage{amsmath,amssymb,amsfonts,graphicx,tensor}
\usepackage{times,palatino,newpxmath,courier}
\usepackage{subcaption}
\newcommand{\dd}{\mathrm{d}}
\usepackage{graphicx}
\usepackage{enumitem}
\usepackage{braket}
\usepackage[dvipsnames]{xcolor}
\usepackage{comment}

\title{Holographic black hole cosmologies}
\author{Abhisek Sahu, Mark Van Raamsdonk}
\affiliation{Department of Physics and Astronomy, University of British Columbia,\\
6224 Agricultural Road, Vancouver, B.C., V6T 1Z1, Canada}

\emailAdd{abhi@phas.ubc.ca}
\emailAdd{mav@phas.ubc.ca}
\date{July 2024}

\abstract{We describe and study a holographic construction of big-bang / big-crunch cosmological spacetimes where the matter consists of a lattice of black holes. The cosmological spacetime is dual to an entangled state of a collection of holographic CFTs associated with the second asymptotic regions of the black holes. For a cosmology with spatial slice geometry $\Sigma$, this state is constructed via a Euclidean path integral for the CFT on a geometry obtained by connecting two copies of $\Sigma$ by a lattice of tubes. In three-dimensional gravity, we describe the cosmological solutions and the associated Euclidean saddles explicitly. For the case of (globally) flat cosmology, we determine when the Euclidean solution associated with the cosmology provides the dominant saddle compared to other natural candidates that preserve the symmetries of the boundary space. We find that the cosmological saddle dominates when the black holes are sufficiently large and close together. 

Our cosmology has a mixed state version where the physics behind the black hole horizons is unspecified and the Euclidean construction involves a pair of CFTs with an ensemble of operator insertions correlated between the two CFTs. Various purifications (adding second asymptotic regions for the black holes) correspond to various ways to promote this ensemble to an interaction by adding auxiliary degrees of freedom that couple the two CFTs in the Euclidean picture. These auxiliary degrees of freedom provide a Hilbert space for the cosmology in the Lorentzian picture.}

\begin{document}

\maketitle
\section{Introduction}

Negative $\Lambda$ gravitational effective field theories associated with holographic CFTs have interesting and potentially realistic cosmological solutions with zero, positive, or negative spatial curvature (for a recent summary, see \cite{VanRaamsdonk:2024sdp}). Understanding whether there is some underlying CFT description of the physics of these solutions may be a straightforward route to achieving fully microscopic models of big-bang cosmologies.

The cosmological solutions are big-bang / big-crunch solutions without any asymptotically AdS regions, so the role of a dual CFT is not obvious. However, in many cases the solutions are time-reflection symmetric at the level of background cosmology and the analytic continuation to Euclidean time is a reflection-symmetric AdS wormhole with two asymptotically AdS boundaries. This suggests that certain special cosmological states may have a Euclidean path integral construction (similar to Hartle-Hawking) where the gravitational path integral has a holographic definition via a pair of holographic Euclidean CFTs \cite{Maldacena:2004rf,VanRaamsdonk:2020tlr}.

The connectedness of the Euclidean wormhole solution suggests correlations between the pair of CFTs and these correlations suggest that some type of ensemble or interaction is present in the holographic construction \cite{Maldacena:2004rf,Saad:2019lba,Betzios:2019rds,VanRaamsdonk:2020tlr}.\footnote{For example, connected two-point functions between the two CFTs computed holographically will be non-vanishing unless there are cancellations due to other saddles in the path integral.} 
Some insight into the nature of these correlations comes by considering a specific simple model of cosmology where the matter is an approximately uniform distribution of heavy particles (associated with CFT operators with large dimension) \cite{MMV} (see also \cite{Chandra:2022bqq, Sahu:2023fbx, Sasieta2022Wormholes, Balasubramanian:2022gmo}).
In the Lorentzian picture, these particles behave as pressureless dust and we get the $\Lambda < 0$ matter cosmology. In the Euclidean solution, the geodesics associated with these particles extend from one asymptotic boundary of the wormhole to the other. This indicates that we have a pair of operator insertions at corresponding points in the two CFTs for each heavy particle.

The operator insertions suggest a natural place for correlations between the CFTs to arise. Instead of considering a single instance of a pair of CFTs with specific operators inserted, we can consider an ensemble of such insertions (in the same pair of CFTs) where the microscopic details of which operators are being inserted vary among elements of the ensemble \cite{Sahu:2023fbx,MMV}. In the cosmology picture, this gives rise to a mixed state cosmology; the ensemble in the CFT description gives rise to an entropy of the cosmological state. For the wormhole to be the dominant saddle in the gravitational path integral, a large degree of correlation is necessary.\footnote{We briefly review an argument from \cite{MMV}: Consider slicing the wormhole along a plane that intersects the two asymptotic boundaries. This slice provides the initial data for a connected Lorentzian wormhole. The dual CFT state here requires ${\cal O}(c)$ mutual information between the two CFTs, and this must arise from the correlations present in the ensemble in the Euclidean picture.} In the cosmology picture, we require a large entropy.\footnote{We will see below that in some cases, we can purify this cosmological state by adding some auxiliary degrees of freedom. In the Euclidean picture, these couple the two CFTs, and the ensemble of insertions arises by integrating them out. In the Lorentzian picture, the auxiliary degrees of freedom provide the Hilbert space in which the state describing the cosmology lives.}

In this paper, we investigate a particular way of achieving this large entropy to produce a cosmological solution that arises from the dominant saddle of a Euclidean path integral. We consider a fixed set of locations for our heavy operator insertions, but for each pair of insertions, we take a specific correlated sum
\begin{equation}
\label{insertions}
    \sum_i {\cal O}^1_i {\cal O}^2_i e^{- \beta \Delta_i } \; .
\end{equation}
For small enough $\beta$, this is dominated by heavy operators associated with black hole microstates and gives the usual canonical ensemble of microstates associated with a thermal CFT state. The resulting cosmology is one where the matter is an array of black holes. 

\begin{figure}
    \centering
    \includegraphics[width=\linewidth]{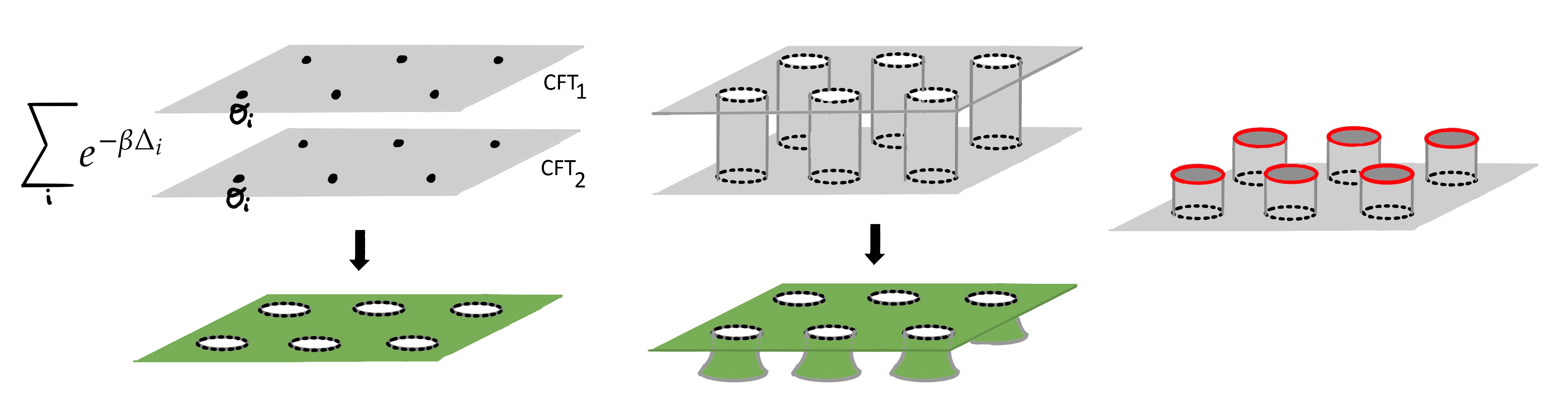}
    \caption{Left: a pair of Euclidean CFTs with an ensemble of operator insertions. Each pair of insertions is summed over all operators with a Boltzmann-type weight. The dual geometry has a spatial slice (green) with a collection of thermal mixed-state black holes. Center: the sum of operator insertions can be obtained by integrating out CFT degrees of freedom on a collection of tubes connecting the two CFTs. The dual geometry in this case has a second asymptotic region for each black hole. Right: The path integral with tubes can be sliced to define a pure entangled state of a collection of non-interacting CFTs (red) on spatial spheres. This state is dual to the Lorentzian cosmology.}
    \label{fig:6pack}
\end{figure}

In our construction, the highly mixed cosmological state can be purified by adding a second asymptotic region to each black hole. In this case, the full cosmology including these second asymptotic regions can be understood as being dual to an entangled state of a collection of CFTs associated with these asymptotic regions.\footnote{From this perspective, the cosmology is the interior geometry of a multi-boundary Lorentzian wormhole. The spatial slice of this wormhole fixed by time-reflection symmetry is depicted schematically in Figure \ref{fig:Bholes}.} In the Euclidean picture, this purification process corresponds to replacing the ensemble with an interaction: each ensemble of insertions (\ref{insertions}) can be obtained by integrating out the CFT degrees of freedom on a tube that connects the two copies of the CFT, as shown in Figure \ref{fig:6pack}. Slicing the resulting path integral in half gives the path integral that constructs the pure entangled state of many CFTs that is dual to the cosmological spacetime (Figure \ref{fig:6pack}, right). 

The entropy of the black holes increases for decreasing $\beta$ in \ref{insertions} or equivalently, as the tubes in the geometrical picture become shorter or wider. We would like to understand whether there is some critical $\beta$ below which the dominant saddle yields a cosmological solution. In order to gain analytic control over this question, we investigate in this paper the case of three dimensional gravity where we are able to write down the black hole cosmology solutions explicitly (following \cite{MMV}). In this case, we do find a critical $\beta$ below which the cosmology dominates; the transition to this cosmological state is akin to the Hawking-Page transition. In the cosmology phase, the black holes are large and close together for the (globally) flat cosmologies that we study in detail. The critical black hole size decreases as the overall volume of the space increases. In the limit of an infinite space with a planar square lattice of black holes, we find that the critical radius is $r_{crit} \approx 1.1 \ell_{AdS}$, approximately equal to the radius $r_{crit} = \ell_{AdS}$ of a 3D black hole in AdS just above the Hawking-Page transition.\footnote{The spacing between the black holes is determined to be $D_{crit} \approx 1.3 \ell_{AdS}$.} Thus, the black hole size is always at least of order the cosmological scale. 

For larger $\beta$ (with smaller and/or more widely separated black holes), we still have cosmological solutions. These correspond to subdominant contributions to the gravitational path integral, but the wavefunction of the universe produced by the CFT construction still contains these cosmologies as rare parts of the superposition.\footnote{It may be worthwhile to keep in mind that our own universe might be a rare part of some, e.g. for anthropic reasons.} Perhaps some projection on the entangled CFT state could isolate the cosmological part of the wavefunction in this case.

\subsubsection*{Outline}

The basic setup we have just described is explained in more detail in section 2 below. 
In section 3, we describe the construction of solutions for the case of three-dimensional gravity. Our approach makes use of the fact that for three-dimensional gravity, solutions involving heavy particles and black holes are locally AdS. The spatial slices of these solutions can be obtained by gluing together pieces of hyperbolic space. For massive particles, these spatial slices include conical singularities. As the particle mass increases above the black hole threshold, these singularities open up into Einstein-Rosen bridges to second asymptotic regions. When the matter is all black holes, the Euclidean gravity solution is smooth and locally AdS everywhere. We explicitly construct such solutions corresponding to arbitrary spherical, planar, or hyperbolic lattices of black holes. For a fixed lattice, we have a one-parameter family of solutions, with the black hole horizon size varying from 0 to infinity. 
In the solutions, the scale factor is always $a(t) = \cos(t/ \ell_{AdS})$ and the volume of space per black hole is always
\begin{equation}
   V/\ell_{AdS}^2 = 2p ( {\frac{\pi}{2}} - {\frac{\pi}{q}}) \; .
\end{equation}
where $p$ is the number of sides of the polygons in the lattice and $q$ is the number of polygons joined at a vertex.

In section 4, we study in detail a particular example where we can control the overall volume of the universe and the size of the black holes. Specifically, we take a square lattice of black holes on a toroidal spatial geometry. In the limit of large volume, we get a spatial geometry that is flat at large scales. Here, we are able to identify and construct three saddles that preserve the lattice symmetry and are the most plausible candidates for the least action saddle at each $\beta$. Using tools developed in previous work \cite{Krasnov2000Holography, Maxfield2016HigherGenus, Wien2017Numerical}, we are able (in Section 4) to compare the actions for these saddles, finding that the cosmological saddle dominates below some critical $\beta$.   

In section 5, we provide further discussion of various points, including how the construction can be generalized to higher dimensions, how the mixed cosmology can be purified in other ways, and how various regions of the cosmological spacetime lie in entanglement wedges of various collections of the dual CFTs. 

\subsubsection*{Relation to earlier work}

Our construction of a black hole filled cosmology provides an example of the type of construction suggested originally in \cite{Maldacena:2004rf} where the cosmology is obtained via analytic continuation from a Euclidean wormhole.  The explicit construction of a wormhole and the associated cosmology based on interactions between a pair of CFTs was discussed in \cite{Cooper2018} and the later related work \cite{Betzios:2019rds,Chen:2020tes,Betzios:2021fnm,VanRaamsdonk:2020tlr,VanRaamsdonk:2021qgv,Sahu:2023fbx}. See also \cite{Usatyuk:2024mzs} for a recent discussion of holographic constructions of negative $\Lambda$ cosmology in low dimensions and \cite{Antonini:2023hdh} for a different construction of closed universe cosmologies making use of auxiliary entangled CFTs. The idea of an ensemble of operator insertions giving rise to a cosmological wormhole was discussed in  \cite{Sahu:2023fbx,MMV}. Lattice black hole cosmologies have been studied by various authors in the past outside the constext of holography; for a recent review, see \cite{bentivegna2018black}.

\section{Setup}

In this section, we describe the setup in detail, starting with the final picture of a connected Euclidean path integral giving rise to a purified cosmology.

Consider a $d$-dimensional holographic CFT, with a corresponding dual $\Lambda < 0$ gravitational effective field theory that describes the physics in the asymptotically AdS spacetimes dual to the CFT states. We describe a Euclidean path integral construction for a state of many such CFTs whose dual gravitational wavefunction includes a description of approximately homogeneous and isotropic big-bang / big-crunch cosmological spacetimes filled with black holes.

Let $B$ be the homogeneous and isotropic geometry (sphere, plane or hyperbolic space) that we want the spatial geometry of our cosmology to approximate. 
We consider two copies of the Euclidean version of our holographic CFT on $B$. Next, we deform the geometry to introduce a number of tubes connecting the two copies of $B$. Topologically, we remove a collection of $n$ $d$-dimensional balls from each copy of $B$ and then glue each $(d-1)$-sphere boundary to the corresponding boundary on the other copy of $B$. We can take the tubes to be spherically symmetric, characterized by some length and radius. We denote by $B_2$ this CFT geometry with two copies of $B$ connected by tubes.

\begin{figure}
    \centering
    \includegraphics[scale = 0.3]{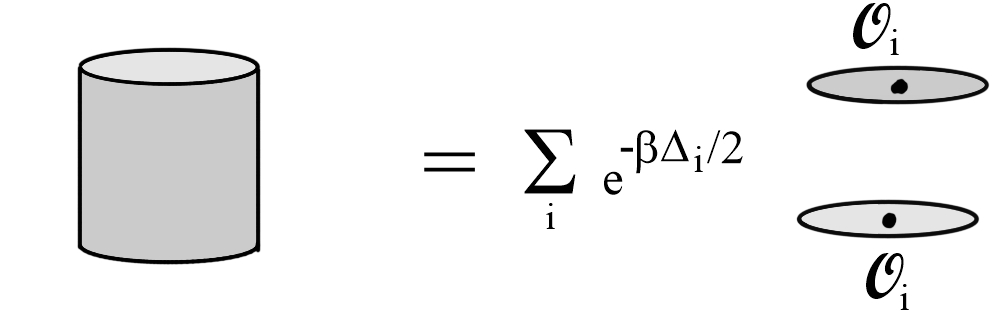}
    \caption{The CFT path integral on a cylinder produces the thermofield double state of the a pair of CFTs on the boundary spheres. The same state is constructed by the Euclidean path integral on a pair of disks with a correlated insertions of operators.}
    \label{fig:TFDtube}
\end{figure}

On its own, the CFT path integral on each cylindrical tube constructs a thermofield double state of a pair of CFTs on the boundary spheres. This state can also be constructed by a Euclidean path integral on a pair of disks with a correlated insertions of operators, as shown in Figure (\ref{fig:TFDtube}). Thus, we expect that for the purpose of evaluating correlators on $B_2$ with no operators inserted on the tubes, it should be equivalent to replace the tubes with correlated insertions of operators on pairs of disks. On the other hand, the tube picture gives us some additional observables (we can insert operators on the tubes), and on the gravity side, gives rise to a more complete geometry.

A key point in either description is that we are introducing correlations between the fields in one copy of the CFT and the fields in the other copy of the CFT. These correlations are reflected in the existence of a Euclidean wormhole solution that connects the two copies of $B$ through the bulk\footnote{Technically, this isn't a usual wormhole since the boundary is connected because of the tubes.}.

The wormhole solution we are interested in is one where the bulk geometry topologically fills the region between the two copies of $B$, leaving the tubes unfilled. This is depicted in \cref{fig:Topologies}. We will refer to this as the ``cosmological saddle'' for reasons that will become apparent. For the same boundary geometry $B_2$, there will be a variety of other possible bulk saddles (i.e. solutions of the bulk gravitational equations whose boundary geometry is conformal to $B_2$). The bulk solution with the least gravitational action will dominate the path integral; later, we will be interested in understanding when the cosmological saddle dominates. This will depend on the parameters of the CFT setup.

\subsubsection*{The Lorentzian geometry}

The Euclidean CFT geometry has a $\mathbb{Z}_2$ reflection symmetry that exchanges the two copies of $B$. The surface fixed by this reflection is a collection of $n$ $(d-1)$-dimensional spheres that comprise the middle slices of all the tubes we have introduced. The Euclidean path integral sliced at this surface constructs an entangled state of $n$ copies of the (Lorentzian) CFT on these spheres. 

The CFT state is dual to a gravitational state that is a superposition of different spacetimes, but is expected to be dominated by one particular geometry that is an analytic continuation of the dominant saddle in the Euclidean construction. According to the standard rules of holography, the dual geometry includes $n$ asymptotically AdS regions with spherical boundary geometry. In some cases, the dominant contribution to the dual geometry might be $n$ copies of pure AdS with matter entangled between the different copies (similar to the dual of the thermofield double state below the Hawking-Page transition). However, we are most interested in the situation where the dominant saddle in the Euclidean construction is the cosmological saddle described above. 

\begin{figure}
    \centering
    \includegraphics[scale = 0.3]{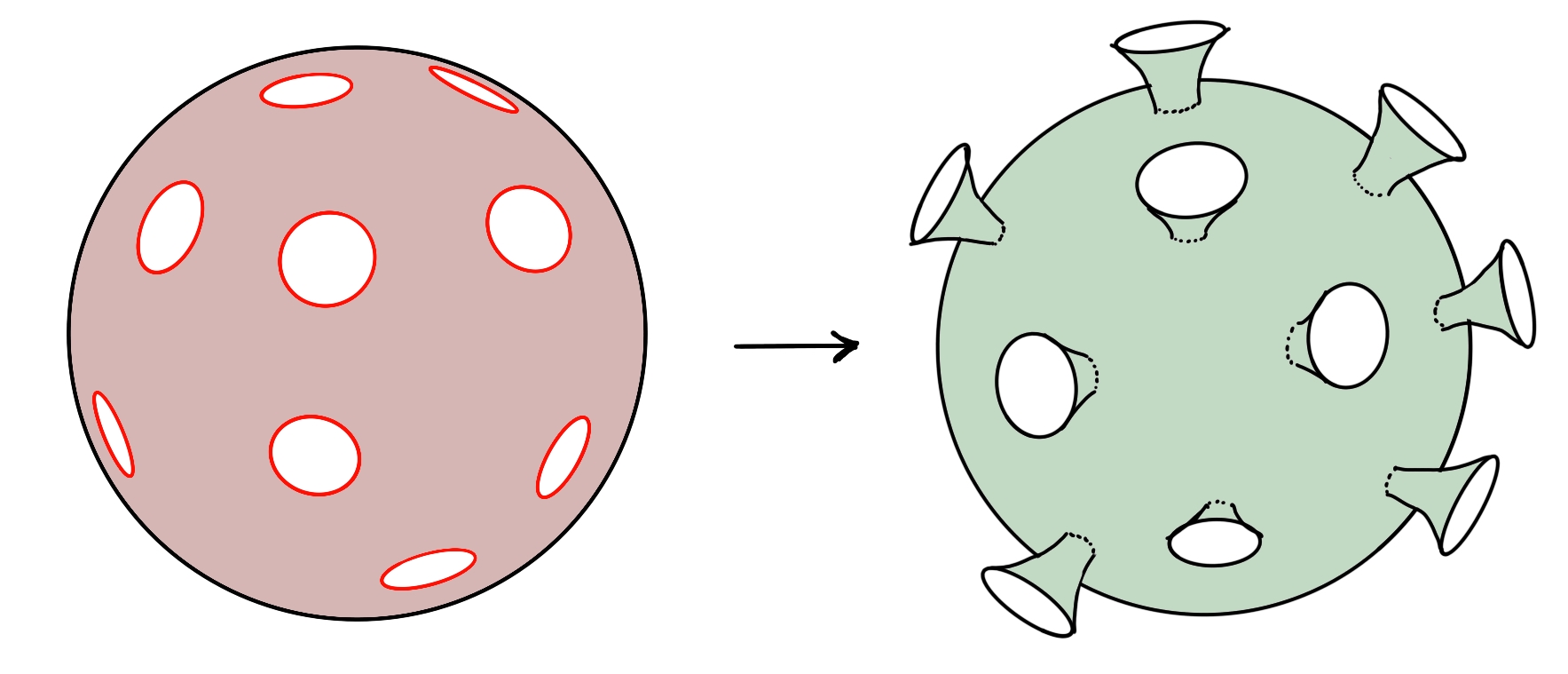}
    \caption{Left: Euclidean path integral constructing an entangled state of CFTs (associated with the red boundaries). Right: Spatial slice of the dual spherical black hole cosmology; the funnels represent the second asymptotic regions for the black holes in the cosmology.}
    \label{fig:Bholes}
\end{figure}

Let us understand the Lorentzian geometry of the cosmological saddle. This background geometry inherits a time reversal symmetry from the Euclidean construction. The spatial geometry of the time-reversal symmetric slice is the same as that in the Euclidean geometry. Since this slice cuts through all of the tubes, the spatial geometry must contain $n$ asymptotic boundaries. But this spatial slice is connected, with topology the same as $B$ with $n$ balls removed. This spatial slice geometry is depicted schematically as in Figure \ref{fig:Bholes} for the case when $B$ is a sphere. We will describe this geometry very explicitly in the case of 3D gravity below.

From each of the asymptotic regions, the spacetime geometry should look like a black hole exterior. However, from the interior perspective, we have a connected geometry with $n$ black holes\footnote{Specifically, we have $n$ spherical codimension-two extremal surfaces on the time-symmetric slice}. We will see below that this interior geometry is time-dependent, with singularities in the past and future, so we can think of it as a big-bang / big-crunch cosmology where the matter is $n$ black holes.

\section{Three dimensional gravity construction}

In this section, we describe the CFT construction and the dual geometries explicitly in the case where we have a two-dimensional holographic CFT and a three-dimensional bulk geometry. 

In three-dimensional gravity, the Einstein equations with negative cosmological constant imply that the spacetime geometry is locally AdS${}_3$ in the absence of matter. The solutions we are interested in have a $\mathbb{Z}_2$ time-reversal symmetry. The surface $\Sigma$ left fixed by this symmetry is an extremal surface whose local geometry is two-dimensional hyperbolic space. Denoting the metric for the geometry of this surface by $d \Sigma^2$, the metric for the full spacetime geometry is simply
\begin{equation}
\label{Lor}
    ds^2 = -dt^2 + \cos^2(t/\ell) d \Sigma^2
\end{equation}
in the Lorentzian picture and 
\begin{equation}
\label{Euc}
    ds^2 = d \tau^2 + \cosh^2(\tau/\ell) d \Sigma^2
\end{equation}
in the Euclidean picture.

\subsection{Spatial geometry}

We can describe the surface $\Sigma$ by gluing together patches of hyperbolic space bounded by geodesics. These patches can be finite area hyperbolic polygons or infinite area patches that include one or more boundary segments on the hyperbolic disk.\footnote{Without loss of generality, we can restrict to hyperbolic triangles or regions bounded by three geodesics that include one boundary segment.} In order to have a smooth geometry, we require that any geodesic segments are glued to other geodesic segments of the same length and that the angles at any vertex add to $2 \pi$. 

\begin{figure}
    \centering
    \includegraphics[scale = 0.2]{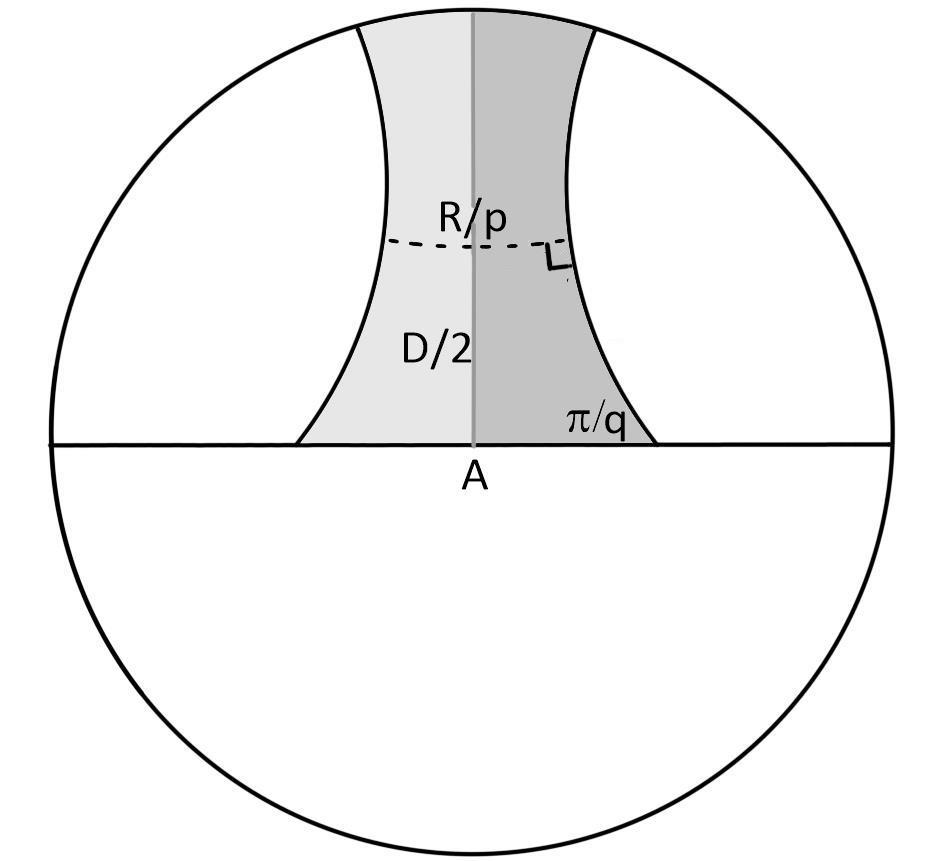}
    \hspace{2 cm}
    \includegraphics[scale = 0.35]{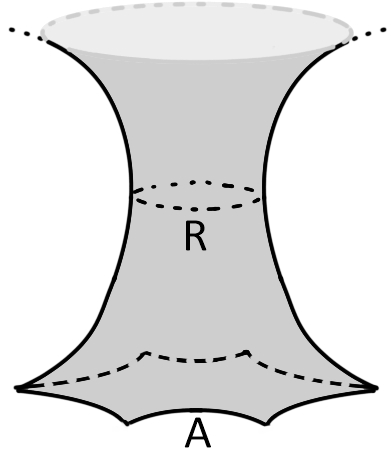}
    \caption{Gluing together patches of hyperbolic disk}
    \label{fig:HypTet}
\end{figure}

\subsection*{Black hole lattices}

As explicit examples, we will consider a geometry that can be understood as a regular lattice of black holes in a space whose global geometry is spherical, flat, or hyperbolic. Cosmologies associated with lattices of black holes have been considered in the past; for a recent review, see \cite{bentivegna2018black}.

We will construct the geometries by gluing copies of a single basic shape shown in figure \ref{fig:HypTet} (left). Here, we have three geodesics on the Poincar\'e disk, with equal intersection angles $\pi/q$ for some $q$ that we will later choose to be an integer. Apart from $q$, we have one other parameter that we can take to be the length $A$ of the finite geodesic segment. By gluing $p$ copies of this shape together along the semi-infinite geodesics, we obtain a funnel geometry shown in Figure \ref{fig:HypTet} (right). The funnel has a geodesic cycle of some length $R$; this will be the black hole horizon length.

For later use, we note that the %distance $D$ in Figure 3 (this will be the distance between the black holes) and the 
two-dimensional volume $V$ (i.e. area) of the space on the finite side of the geodesic cycle is given by
%\begin{equation}
%\left(\sinh^2{D \over 2} + {1 \over \cosh^2{R \over 2 p}}\right)\left(\sinh^2{R \over 2p} + {1 \over \cosh^2{D \over 2 }}\right) = \left(1 + \sinh{R \over 2p} \sinh{D \over 2} \cos {\pi \over q} \right)^2
%\end{equation}
%and 
\begin{equation}
\label{volume}
V/\ell_{AdS}^2 =  2 p \left({\pi \over 2} - {\pi \over q} \right) 
\end{equation}
as a consequence of the fact that the area of an $n$-sided polygon in hyperbolic space is equal to $\ell_{AdS}^2$ times the amount by which the sum of the interior angles is less than $(n-2)\pi$. 

We now form a lattice of these funnels by gluing the polygonal boundaries together, with $q > 2$
\footnote{For $q=2$ we get the BTZ black hole where two infinite funnels join at a codimension 1 circle, and there is no cosmological region. }
polygons joined at each vertex. Since the vertex angles of each funnel are $2 \pi / q$, the resulting geometry is smooth everywhere. For various cases of $(p,q)$, we can obtain spherical, planar, or hyperbolic lattices
\footnote{The Euler characteristic $\chi$ of such a lattice is $pF\left(\frac{1}{p}+\frac{1}{q}-\frac{1}{2}\right)$, assuming there are no boundaries. We have a hyperbolic lattice for $\chi<0$, a planar lattice for $\chi=0$ and a spherical lattice for $\chi>0.$}
, and in the last two cases, these can be infinite or (by some identifications) finite.

\subsubsection*{Spherical lattices}

For the cases $(p,q) \in \{(3,3),(3,4),(3,5),(4,3),(5,3)\}$, we have a spherical lattice of black holes arranged in a regular tetrahedron, octahedron, icosahedron, cube, and dodecahedron respectively. From \ref{volume}, the two-dimensional volume $V$ of of the space outside the horizons are related to $R$ by
\begin{equation}
V/\ell_{AdS}^2 =  2 p F \left({\pi \over 2} - {\pi \over q} \right) 
\end{equation} where $F$ is the number of faces of the polyhedron.

\subsubsection*{Planar lattices}

The cases $(p,q) = \{(3,6),(4,4),(6,3)\}$ correspond to planar triangular, square, and hexagonal lattices. These can be taken as infinite, or, by making appropriate identifications, toroidal. For our analysis below, we will consider in detail the case where we have an $m \times n$ square lattice.

\subsubsection*{Hyperbolic lattices}

For any other $(p,q)$ with $q > 2 + 4/(p-2)$, we have a hyperbolic lattice. As for the planar case, this can be taken to be infinite, or, by various identifications, a finite lattice of genus $g > 2$. 

\subsection{Euclidean spacetime geometry}

\begin{figure}
    \centering
    \includegraphics[width=\linewidth]{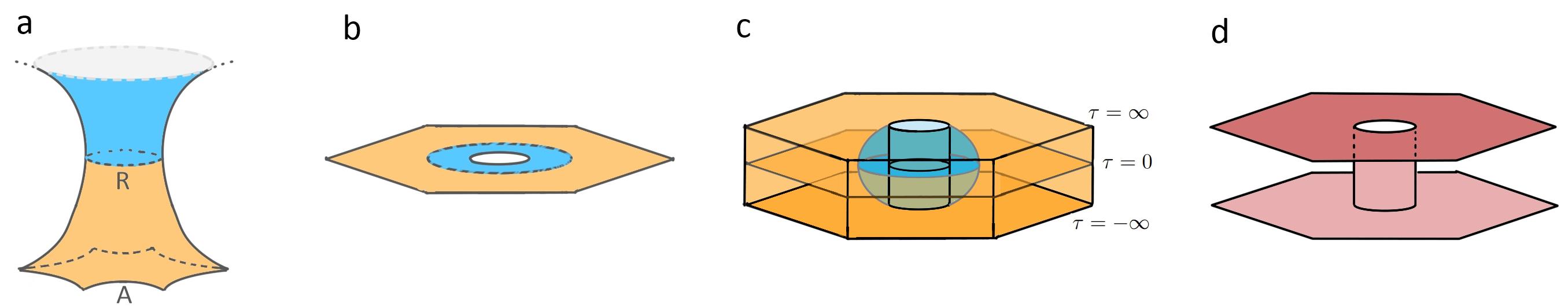}
    \caption{a) The hyperbolic funnel geometry of the $\tau = 0$ slice. b) a simpler representation of this, with the asymptotic boundary represented as a finite circle. c) The spacetime geometry (\ref{Euc}) corresponding to this spatial patch. d) The asymptotic boundary of this spacetime geometry. }
    \label{fig:EucST}
\end{figure}

Let us now understand better the Euclidean spatial geometry (\ref{Euc}) arising from the spatial slices that we have described. The three dimensional geometry can be constructed by gluing together patches described by (\ref{Euc}) where $d \Sigma^2$ is the hyperbolic funnel shown in Figure \ref{fig:HypTet} (right). For each funnel, the part on the finite side of the geodesic cycle is topologically an annulus and gives rise to a 3D geometry that topologically is a thickened version of this annulus with two annular boundaries. The funnel on the infinite side of the geodesic cycle gives rise to a spacetime geometry that is half a Euclidean BTZ black hole. This has a boundary geometry that is a finite cylinder which connects the two asymptotic boundaries on the annular part, as shown in Figure \ref{fig:EucST}c. Thus, the spacetime geometry arising from the full funnel has a connected boundary( Figure \ref{fig:EucST}d).

For all the cases we have described above, the full spacetime geometry is obtained by gluing together a lattice of these three-dimensional pieces in the same way that we glued together the two-dimensional pieces to obtain the spatial slice. The boundary geometry of the resulting space is connected, but with some large genus; see Figure \ref{fig:LatticeBoundary} for the case of a flat toroidal boundary. For the spherical lattices with $n$ black holes, the genus of this boundary surfaces is $n-1$. 

\subsection{Lorentzian spacetime geometry}

In the Lorentzian version (\ref{Lor}) of the geometry, the infinite side of each funnel leads to a spacetime region that is equivalent to one half of a two-sided BTZ black hole (including the exterior region and half of the past and future interior regions). The finite interior region leads to the cosmological part of the spacetime with big-bang and big-crunch singularities in the past and future and a lattice of black holes. The spatial slices of this cosmological region are locally negatively curved, but the geometry on large distance scales can be spherical, planar, or hyperbolic, depending on which type of lattice we choose.  

\subsection{Other saddles}

We have described a variety of solutions that from the exterior perspective are connected multi-boundary wormholes and from the interior perspective are big-bang / big-crunch cosmologies filled with a lattice of black holes.

In each case, the associated Euclidean solution obtained by analytic continuation has a connected two-dimensional boundary geometry. In general, there will be other gravitational solutions with this same boundary geometry. In order that the CFT path integral on this boundary geometry produces a state that is dominated by the cosmology, the cosmological saddle that we have described must be the least action solution with the given boundary geometry. In the next section, we will investigate in detail for a specific case when this cosmological saddle dominates.

\section{Comparing actions}

In order that the cosmological solution is the dominant contribution to the wavefunction, it must be that the 3D Euclidean gravity solution corresponding to the cosmology is the least action solution with the given boundary geometry. Whether or not this is true will depend on the parameters of the solution, namely the number and arrangement of black holes and their horizon radii. In this section, we will investigate when the cosmological saddle dominates in a specific case.

\begin{figure}
    \centering
    \includegraphics[scale = 0.20]{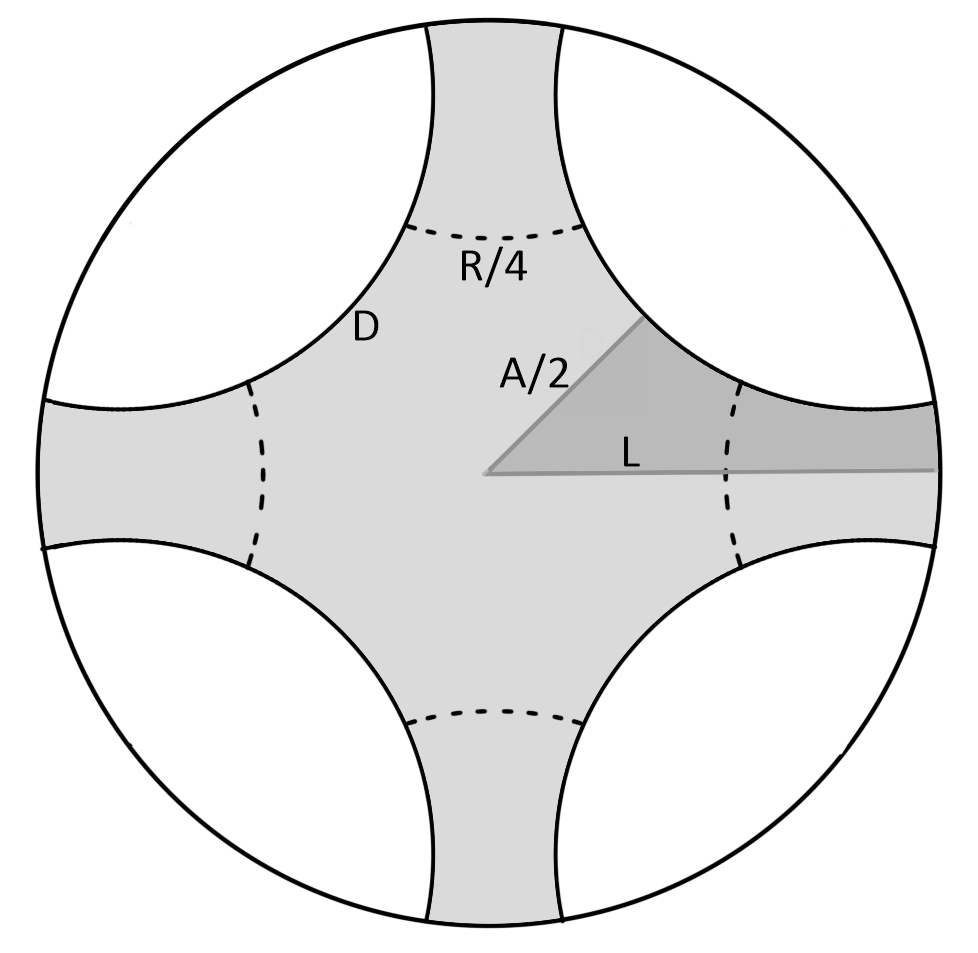}
    \caption{Hyperbolic patch for construction of the spatial geometry with a square lattice of black holes. The shaded region here corresponds to the shaded region in Figure \ref{fig:HypTet}a. The full geometry is obtained by gluing together an $M \times N$ array of copies of this this along the geodesic sides.}
    \label{fig:SquareLat}
\end{figure}

We would like to consider a class of solutions where we can have an arbitrary number of black holes with an arbitrary horizon radius. It will be convenient in our calculations to consider the situation where we have a planar lattice with identifications to give an $m \times n$ square lattice of black holes, with each lattice ``site'' preserving $D_4$ symmetry. In this case, geometry of the $t=0$ spatial slice of the cosmological saddle can be obtained by gluing together an $m \times n$ grid of hyperbolic funnels with square boundaries. Alternatively, we can glue together an $m \times n$ grid of the geometry shown in Figure \ref{fig:SquareLat}. 
Apart from $m$ and $n$, we have only a single physical dimensionless parameter in the setup, which can be taken to be the horizon length $R$ of one of the black holes in the cosmological saddle for the given boundary geometry.

\begin{figure}
    \centering
    \includegraphics[width=0.7\linewidth]{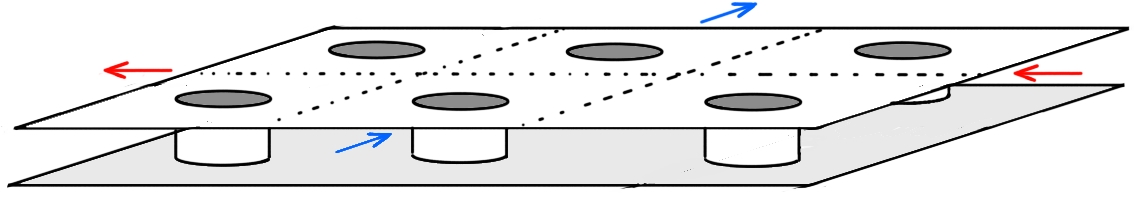}
    \caption{Euclidean boundary geometry for cosmological saddle with toriodal spatial geometry and and a lattice of six black holes.}
    \label{fig:LatticeBoundary}
\end{figure}

The full boundary geometry for this solution is shown schematically in Figure \ref{fig:LatticeBoundary}. This has genus $(mn + 1)$.
The Euclidean cosmological saddle can be obtained by gluing together an $m \times n$ lattice of 3D pieces, as we have described above. 
For this solution, the bulk geometry topologically fills in the region between the planes in Figure \ref{fig:LatticeBoundary} but does not fill in the tubes. However, there are other solutions that preserve this lattice symmetry and which have lower action in some cases.

To understand these other solutions, note the boundary geometry of Figure \ref{fig:LatticeBoundary} is topologically a pair of tori glued together via a lattice of tubes. Other solutions preserving the lattice symmetry correspond (topologically) to filling in these tori in some way and then gluing the filled tori together with solid tubes. As an example, the topologies of three different saddles are shown in Figure \ref{fig:Topologies} for the $1 \times 2$ lattice.

\subsection{Simplification using lattice symmetry}

We would now like to set up the comparison of gravitational actions for these simplest bulk saddles preserving the full symmetry of the boundary geometry. Let $M_{m,n}$ be the boundary geometry in some conformal frame.\footnote{Later, it will be convenient to consider the frame in which this geometry has constant negative curvature.} 

\begin{figure}
    \centering
    \includegraphics[scale = 0.3]{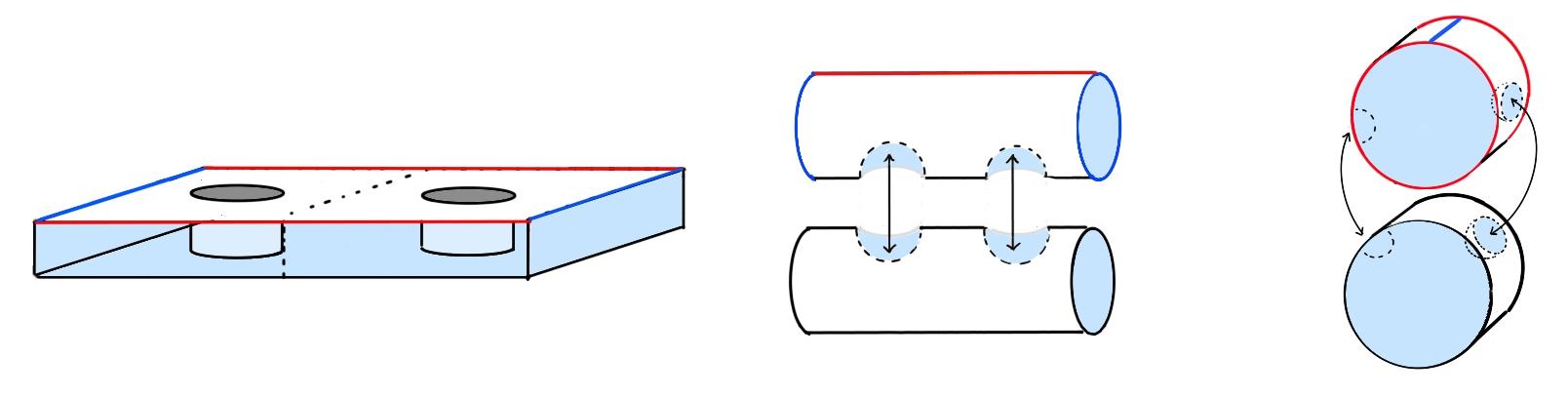}
    \caption{Topological description of bulk saddles for the same boundary geometry. The left most panel shows the cosmological saddle whereas the other two show the non-cosmological ones. Red edges and blue edges are identified.}  
    \label{fig:Topologies}
\end{figure}

We will consider the three saddles with topology shown in Figure \ref{fig:Topologies}, referring to these as saddle 1,2, and 3 respectively. Let $S^{(i)}_{m,n}$ be the action (in some conformal frame) for the $i$th saddle with boundary geometry $M_{m,n}$. For the cosmological saddle, the bulk geometry is just an $m \times n$ grid of identical pieces, and each of these pieces will have the same action as the cosmological saddle for $M_{1,1}$. Thus, we have
\begin{equation}
    S^{(1)}_{m,n} = m n S^{(1)}_{1,1} \; .
\end{equation}
Saddles 2 and 3 can be divided into $m$ and $n$ identical pieces respectively, and the action for each of these pieces will be the same as the action for $M_{1,n}$ and $M_{m,1}$ respectively, so we have that
\begin{equation}
    S^{(2)}_{m,n} = m  S^{(1)}_{1,n} \qquad \qquad S^{(3)}_{m,n} = n  S^{(3)}_{m,1} \; .
\end{equation}
Using these, we can write the action differences as
\begin{eqnarray*}
    S^{(1)}_{m,n} - S^{(2)}_{m,n} &=& mn  \left[ (S^{(1)}_{1,1} -  S^{(2)}_{1,1}) + (S^{(2)}_{1,1} - {1 \over n} S^{(2)}_{1,n})\right] \cr
    S^{(1)}_{m,n} - S^{(3)}_{m,n} &=& mn  \left[ (S^{(1)}_{1,1} -  S^{(3)}_{1,1}) + (S^{(3)}_{1,1} - {1 \over m} S^{(3)}_{m,1})\right]
\end{eqnarray*}
The cosmological saddle will dominate if both of these are negative. 

For simplicity, we will focus on the case where $m=n$ where saddles 2 and 3 are related by the symmetry that exchanges rows and columns of the lattice. Thus, we would like to understand when the expression 
\begin{equation}
\label{eq:nncomp}
    S^{(1)}_{n,n} - S^{(2)}_{n,n} = n^2  \left[ (S^{(1)}_{1,1} -  S^{(2)}_{1,1}) + (S^{(2)}_{1,1} - {1 \over n} S^{(2)}_{1,n})\right]
\end{equation}
is negative. 
Here, the first difference in brackets is the action difference when the boundary geometry is the genus two surface $M_{1,1}$. In this case, the cosmological saddle corresponds to a toroidal universe with a single black hole. In section \cref{subsec:Genus2term} below, we will compute this action difference as a function of the parameter $R$ that gives the horizon length of the black hole, finding that it decreases with $R$ and is negative for $R$ above some critical size $R^{crit}_1$. Above this critical size, the cosmological saddle dominates in the single black hole toroidal universe.

The second difference becomes relevant for $n > 1$. In \cref{subsec:Replica} below, we will argue that for fixed $R$ this contribution is negative and decreases with $n$. Thus, with more black holes, the critical black hole size can be smaller.
Defining the asymptotic value 
\begin{equation}
    \Delta(R) = \lim_{n \to \infty} \left({1 \over n}S^{(2)}_{1,n}(R) - S^{(2)}_{1,1}(R)\right) \; ,
\end{equation}
the critical black hole size $R^{crit}_\infty$ in the limit of an infinite grid will be determined as
\begin{equation}
     (S^{(1)}_{1,1} -  S^{(2)}_{1,1})(R) = \Delta(R)  \; .
\end{equation}

\subsection{Action difference for genus two}
\label{subsec:Genus2term}
In this section, we consider the first term in \cref{eq:nncomp}, that is, the action difference $(S^{(1)}_{1,1} - S^{(2)}_{1,1})$. Each of the actions appearing here is the gravitational action for a different bulk geometry with the same genus two boundary. 

We will consider a one-parameter family of boundary geometries parameterised by the size of the black hole horizon in the cosmological saddle. To describe this, it is useful to start with the time-symmetric slice of the bulk cosmological solution, obtained by making identifications on the Poincar\'e disk shown in Figure \ref{fig:SquareLat}. 

A useful description of the full 3D geometry of the cosmological saddle can be obtained by taking the Poincar\'e disk in Figure \ref{fig:SquareLat} as the surface $x^2+y^2+z^2 = 1$ in the Poincar\'e-Ads geometry with metric
\begin{equation}
    ds^2 = {\ell^2 \over z^2} (dz^2 + dx^2 + dy^2) \; .
\end{equation}

The cosmological saddle geometry is obtained by 
by removing half balls of radius $ \rho < 1$ centred at $(\pm \sqrt{(1 + \rho^2)/2}, \pm \sqrt{(1+\rho^2)/2}$ and identifying the boundaries of the half-balls on opposite corners of the square. This identification is smooth since each boundary introduced is a geodesic $H_2$ surface in $H_3$. In addition to the obvious symmetries inherited from the construction, this geometry has a symmetry 
\begin{equation}
    z \to {z \over x^2 + y^2 + z^2} \qquad x \to {x \over x^2 + y^2 + z^2} \qquad y \to {y \over x^2 + y^2 + z^2}
\end{equation}
that corresponds to the $\mathbb{Z}_2$ Euclidean time reflection. 

The bulk geometrical parameters $R$ (the black hole horizon length) and $D$ (the distance between black holes) shown in Figure \ref{fig:SquareLat} are related to the parameter $\rho$ by
\begin{equation}
\label{DvsR}
    \tanh {D \over 2}  = \rho \qquad \qquad \cosh {R \over 4} = {1 \over \rho^2} \; .
\end{equation}
To obtain these, we note that the sphere $x^2 + y^2 + z^2 = 1$ intersects the four spheres defined above at points with $|x|=|y| = 1/\sqrt{2(1+\rho^2)}$. The distance $A/2$ in \ref{fig:SquareLat} is the geodesic distance between $(x,y,z) = (0,0,1)$ and one of these points. The other results can be obtained from the results of \cref{app:pentagon} by noting that we have a right-angled hyperbolic pentagon with sides $D/2,A/2,A/2,D/2,R/4$.

In this description, the boundary geometry can be described as the plane with disks of radius  $\rho$ centred at $(\pm \sqrt{(1 + \rho^2)/2}, \pm \sqrt{(1+\rho^2)/2}$ removed and the boundaries of diagonally opposite disks identified. This is an example of a more general construction of genus two surfaces known as the Schottky construction. Since the Schottky construction will be useful in describing other saddles with the same boundary geometry, we briefly describe it here.

\subsubsection*{The Schottky description of genus two surfaces}

\begin{figure}
    \centering
    \includegraphics[scale = 0.3]{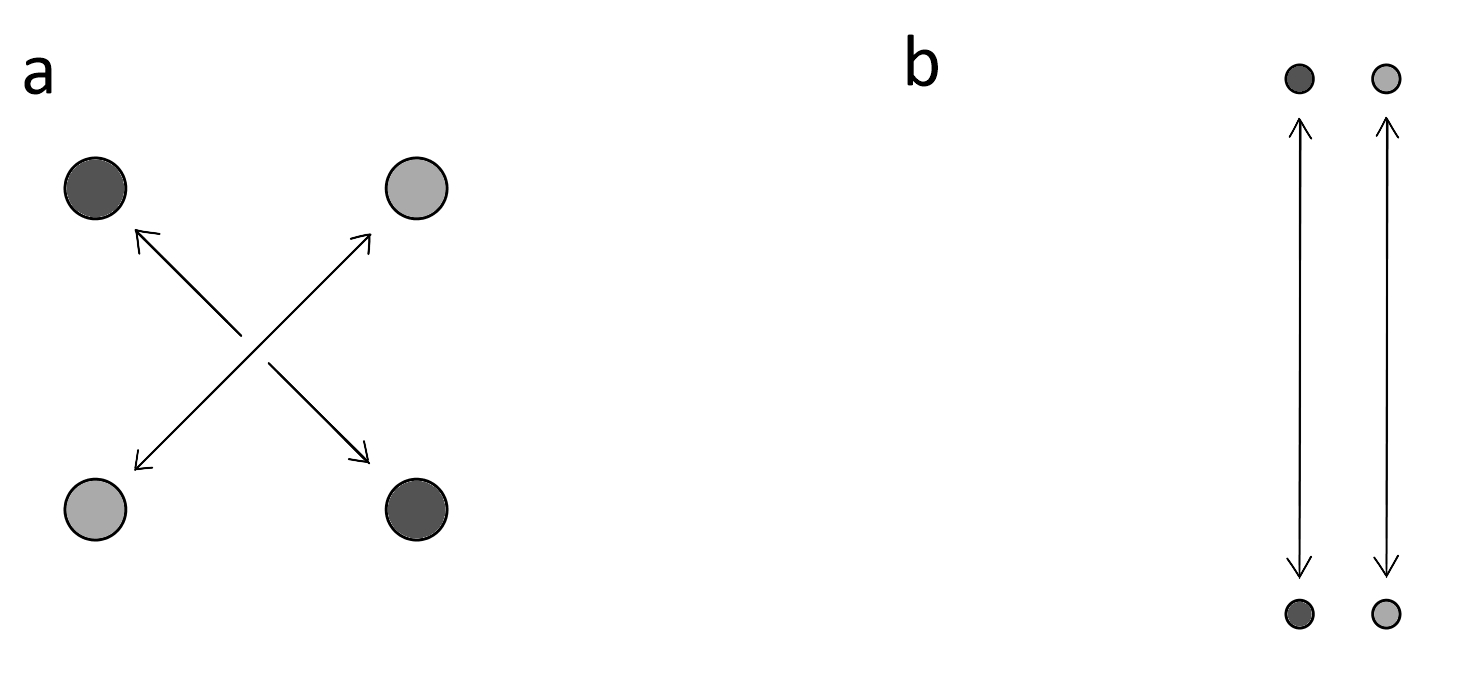}
    \caption{Two different Schottky representations of the genus 2 Bolza curve. For each one, we obtain a corresponding bulk solution by considering the plane as the boundary of Poincar\'e-AdS, removing half-balls corresponding to each circle, and gluing as indicated. The description a gives the cosmological saddle, while the description b gives an alternative saddle.}  
    \label{fig:Bolza}
\end{figure}

The Schottky description of a genus two surface involves two pairs of circles $(C_1,C_1')$, $(C_2,C_2')$ on the complex plane and a pair of elements $L_i$ of $SL(2,\mathbb{Z})$ mapping the exterior of $C_i$ to the interior of $C_i'$. Letting $\Omega_0$ be the region exterior to the circles, and $\Omega = \Omega_0 \cup C_1 \cup C_2$, the genus 2 surface can be understood as a quotient $G \Omega / G$ where $G$ is the group generated by $\{L_1,L_2\}$. The region $\Omega$ is a fundamental domain for the action of $G$ here. The Riemann surface inherits a complex structure from that of the plane. This defines a conformal class of metrics; we will describe the construction of a particular metric with constant negative curvature below. Any genus two Riemann surface can be constructed in this way, but the construction is not unique.

The Schottky description is particularly useful in the holographic context, since given a Schottky construction of some genus two surface, there is a natural 3D locally AdS geometry with this surface as the boundary geometry. To construct this, we consider the Schottky plane as the boundary of Poincar\'e-AdS, remove bulk hemispheres corresponding to each of the circles $C_i$ and $C_i'$, and glue the hemisphere boundary associated with $C_i$ to that associated with $C_i'$. This produces a smooth geometry. 
Different saddles for the same boundary geometry arise as the 3D geometries associated with different Schottky constructions of this boundary geometry.\footnote{These constructions should be non-equivalent under the equivalence relation $L_i \to M L_i M^{-1}$ for $M$ an element of $SL(2,\mathbb{Z})$.}

Our cosmological saddle is associated with a Schottky construction as shown in Figure \ref{fig:Bolza}a, where $L_1$ and $L_2$ map the lower circles to the diagonally opposite upper circles. The circles have radius $\rho$ and are centered at $(\pm \sqrt{(1+\rho^2)/2)}, \pm \sqrt{(1+\rho^2)/2})$. The associated 3D geometry has a $\mathbb{D}_4$ symmetry. 
To find the non-cosmological saddle, we can consider a separate family of Schottky constructions indicated in Figure \ref{fig:Bolza}b. Here, we have unit circles centered at ${u,\bar{u},-u,-\bar{u}}$ where $u = \alpha + i \cosh \lambda$. The generator $L_1$ maps the exterior of the bottom right circle in Figure \ref{fig:Bolza}b to the interior of the top right circle and $L_2$ maps the exterior of the bottom left circle to the interior of the top left circle. 
We describe the above Schottky constructions in more detail in \cref{app:Schottky}.

This construction gives a two-parameter family of genus two Riemann surfaces with $\mathbb{Z}_2 \times \mathbb{Z}_2 \times \mathbb{Z}_2$ symmetry\footnote{There are sub-families of Riemann surfaces which admit larger automorphism groups with 16, 24 or 48 fold symmetry. See \cref{app:Automorphism} for a review.}. There is a one parameter sub-family with $\mathbb{D}_4$ symmetry; for each $\rho$ in the construction associated with the cosmological saddle, there will be some $u(\rho)$ for which the non-cosmological saddle that we have just described has the same boundary geometry. We will explain how to determine $u(\rho)$ presently.

\subsubsection*{Matching boundary geometries}

\begin{figure}[h]
    \centering
    \begin{subfigure}[b]{0.45\textwidth}
        \centering   \includegraphics[width=\textwidth]{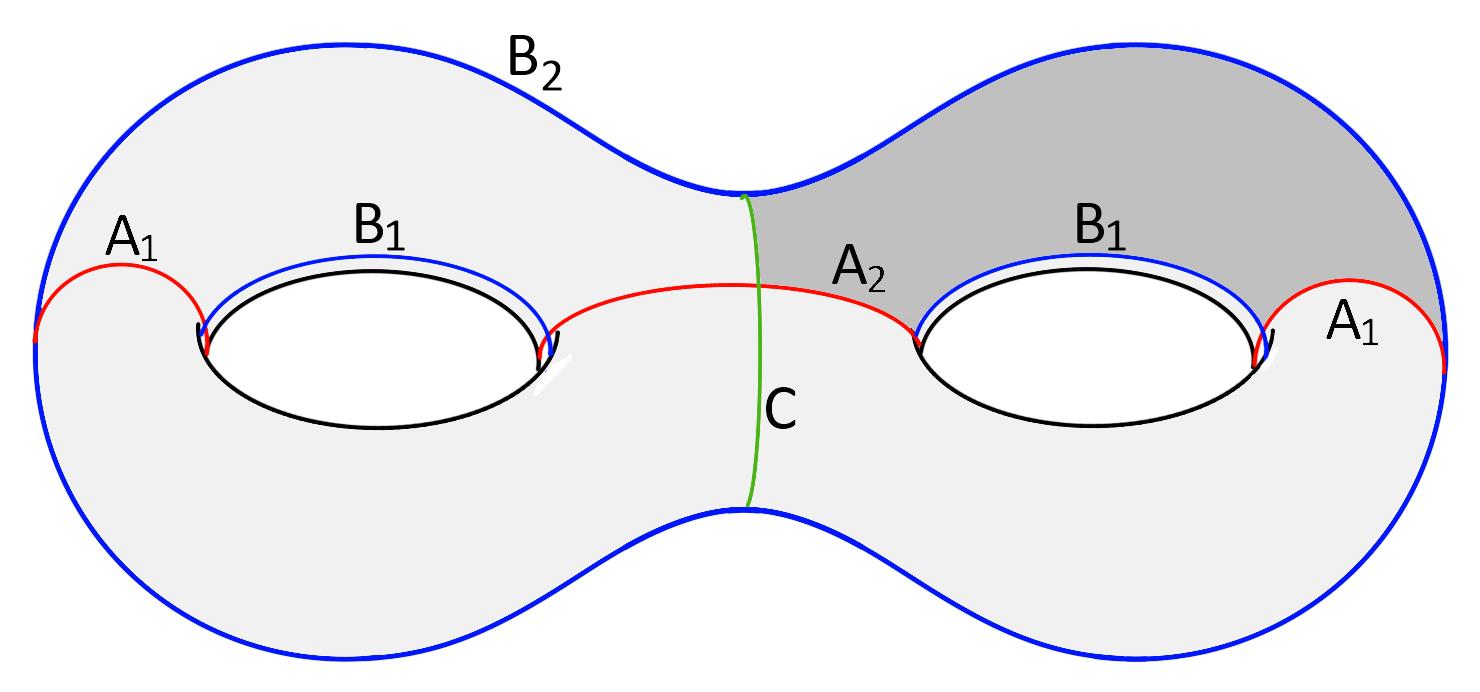}
        \label{fig:image1}
    \end{subfigure}
    \begin{subfigure}[b]{0.45\textwidth}
        \centering
\includegraphics[width=0.6 \textwidth]{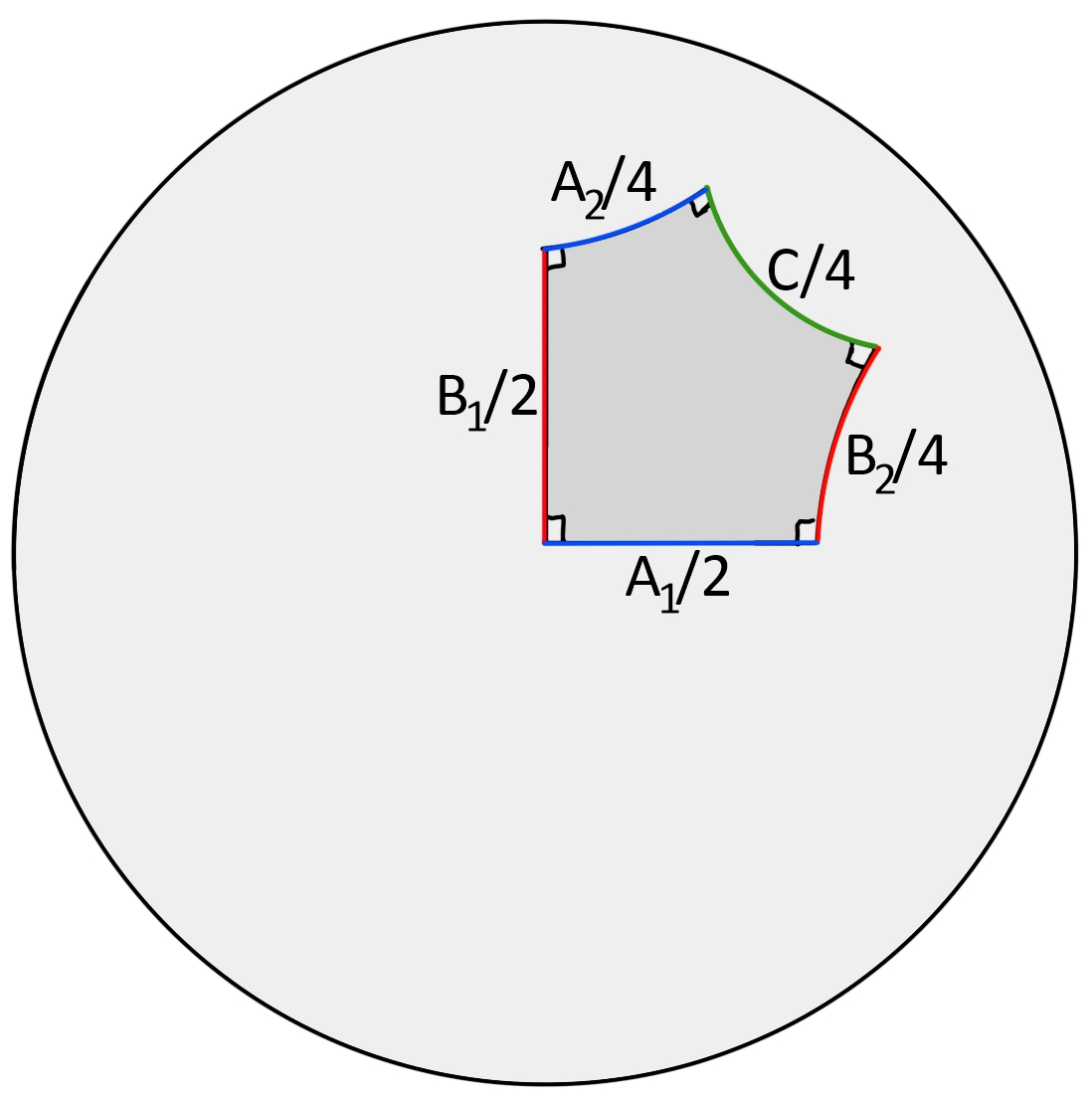}
        \label{fig:image2}
    \end{subfigure}
    \caption{Left: A genus 2 Riemann surface with $\mathbb{Z}_2 \times \mathbb{Z}_2 \times \mathbb{Z}_2$, showing the various geodesics fixed by one of the $\mathbb{Z}_2$ reflections. Right: a right-angled pentagon on the hyperbolic disk. Gluing eight of these pentagons together gives the genus two Riemann surface with a metric of constant negative curvature.}
    \label{fig:Riemann2}
\end{figure}

To find cosmological and non-cosmological saddles with the same boundary geometry, it will be useful to make use of the uniformisation theorem, which guarantees that for any conformal class of genus two surfaces (with elements related by Weyl transformations) there exists a unique geometry with constant negative curvature. The constant curvature metrics we are interested in sit within a two-parameter family with $\mathbb{Z}_2 \times \mathbb{Z}_2 \times \mathbb{Z}_2$ symmetry that can be constructed by gluing eight right-angle hyperbolic pentagons together as shown in Figure \ref{fig:Riemann2}. The pentagon boundaries become geodesics, each fixed by one of the $\mathbb{Z}_2$ symmetries. Using hyperbolic trigonometry, it is straightforward to show that only two of the geodesic lengths are independent, with the remaining ones determined by relations described in Appendix \ref{app:pentagon}. 

We can choose the independent cycle lengths to be $A_1$ and $B_1$. 
The Schottky construction associated with the cosmological saddle gives a one-parameter family of boundary geometries with the further constraint that the cycle lengths $A_1$ and $B_1$ are equal; this gives the additional symmetry that gives us a square lattice of black holes instead of a rectangular lattice. For each value the parameter $\rho$, we have some parameters $A_1(\rho) = B_1(\rho)$ for the boundary Riemann surface in the constant curvature description. 

The Schottky construction associated with the non-cosmological saddle gives a two parameter family of boundary geometries. For each $u$ we have some $A_1(u)$ and $B_1(u)$. These are generally not equal but will be equal in a one-parameter family. In this family, each boundary surface will be equivalent to the surface obtained in the earlier construction for some $\rho$. 

To summarize, for each parameter $\rho$ we can use a numerical method described below to find $A_1(\rho) = B_1(\rho)$. Similarly, for each $u$, we can numerically find $A_1(u)$ and $B_1(u)$, and by scanning over parameters find a $u$ that gives the same values as for a given $\rho$ parameter. We can then compare the actions for the saddles associated with $\rho$ and $u(\rho)$.

\subsubsection*{Numerical computation}

In this section, following \cite{Maxfield2016HigherGenus, Wien2017Numerical}, we describe the numerical procedure to start from a genus two Riemann surface specified by a Schottky presentation and determine cycle lengths on the constant curvature Riemann surface obtained from this via a Weyl transformation.  
With this procedure, we will identify different Schottky constructions that give the same constant curvature Riemann surface. Each Schottky construction has an associated bulk solution that we have described above, so we obtain different bulk saddles for the same boundary surface. Below, we describe how to compare actions for these saddles.

Starting with the flat metric on the plane, the Weyl-rescaled metric with constant negative curvature can be written as 
\begin{equation}
    ds^2 = e^{2\Phi(z,\bar{z})}dzd\Bar{z}.
\end{equation}
The requirement of constant negative curvature implies that the conformal factor $\Phi$ satisfies the Liouville equation 
\begin{equation}
\label{eq:Liouville}
    4\partial_{z}\partial_{\bar{z}} \Phi = %{1\over2}
    e^{2\Phi}.
\end{equation}
To ensure that metric is consistent between various images of the fundamental domain, the Liouville field is subject to the conditions
\begin{equation}
    \Phi(L(z)) = \Phi(z)-{1\over 2}\log{|L'(z)|^2},
\end{equation}
where $L \in \{L_1,L_2\}$.

Analytical solutions of the Liouville equation can be found only in a few highly symmetrical cases. For our calculations, we make use of a Mathematica package \textit{Handlebodies.m}\cite{Wien2017Numerical} developed by Jason Wein. The package takes a description of the circles $C_i,C_i'$ as an input, generates a mesh on the domain $\Omega$ (or a reduced domain if there are symmetries specified) and solves the Liouville equation using finite element methods. The output includes the numerical solution of $\Phi$ and other boundary quantities of interest such as the length of various boundary geodesics.

With this method, we can calculate the lengths $A_1(\rho) = B_1(\rho)$ of the geodesic cycles shown in Figure \ref{fig:Riemann2} starting with the Schottky description associated with the cosmological saddle boundary (Figure \ref{fig:Bolza}a).

Next, we can consider the Schottky description of Figure \ref{fig:Bolza}b, numerically finding $A_1$ and $B_1$ given parameters $\alpha$ and $\lambda$. We scan over the possible values of $\alpha$ to obtain the values of $\lambda$ such that the resulting boundary geometry has $A_1 = B_1$. Finally, by varying $\rho$ or $\alpha$, we can find parameters $\rho$ and $(\alpha,\lambda)$ that each give the same value of $A_1 = B_1$.
Both these searches involve finding a single parameter and can be accomplished using Newton-Raphson or Secant methods.
The result is that we have a one-parameter family of values $(\rho,\alpha,\lambda,A_1,B_1)$ corresponding to the one-parameter family of boundary genus two Riemann surfaces with our desired symmetry.

 In this one-parameter family, as $\alpha$ is increased, $\lambda$ decreases and approaches a constant value of $\pi$ as $\alpha\rightarrow\infty$. The length of the $A_1$ cycle \footnote{Recall that we are considering the one-parameter regime where the A cycles and the B-cycles have equal length, i.e., $A_1=B_1$ and $A_2=B_2$.} (the 2 A cycles with equal length due to one of the $\mathbb{Z}_2$ symmetries) decreases monotonically with increasing $\alpha$ and the length of the $A_2$ cycle (the non-repeating A cycle) increases monotonically with $\alpha$. 
 Consequently, there is a point where the $A_1$ cycle and $A_2$ cycle have equal length and hence all the $A$ and $B$ cycles have the same length $2\cosh^{-1}{2}$. This corresponds to the maximally symmetric genus 2 Riemann surface, also known as the Bolza curve. 
 The Schottky parameter values corresponding to this curve are, $ \alpha=3.178, \lambda = 3.652, \rho = 0.135$. The Schottky descriptions corresponding to these parameters are shown in Figure \ref{fig:Bolza}. 

 In \cite{Hidalgo2005Numerical} the authors use a numerical approach called Burnside's program and find the Schottky parameters corresponding to the Bolza curve.
 The one parameter family of Schottky groups used in their paper, which makes the $\mathbb{D}_6$ automorphism manifest, is different than the one mentioned here which manifests the $\mathbb{D}_4$ automorphism group. 
 However, an appropriate Mobius transformation can be constructed that maps their Schottky group into the one shown in (\cref{fig:Bolza},right). 
 We verified that the Schottky parameters for the Bolza curve match in both our cases under the appropriate Mobius transformation. This provides a check of our numerical procedure.

\subsubsection*{Action calculation}

The Schottky construction based on the parameter $\rho$ and the one based on the corresponding parameters $(\alpha(\rho),\lambda(\rho))$ each have an associated bulk solution, obtained by identifying hemispheres in Poincar\'e-AdS associated with the various circles. Our goal is to compute the action difference between the cosmological saddle (associated with the construction in Figure \ref{fig:Bolza}a) and the other saddle (associated with the construction in Figure \ref{fig:Bolza}b). The action comparison is also accomplished via the \textit{Handlebodies.m}\cite{Wien2017Numerical} package.

The bulk gravitational action is given by
\begin{align}
    S   &= S_{bulk} + S_{GHY} + S_{ct} \\
        &= -\frac{1}{16\pi G}\int d^3x \sqrt{g}(R-2\Lambda) -\frac{1}{8\pi G}\int d^2x \sqrt{h}(K-{1\over \ell}).
\end{align}
The bulk term and the boundary term both diverge near the boundary. Hence one must regularise the action appropriately and cancel the divergent terms before removing the regulator.  
Krasnov \cite{Krasnov2000Holography,Krasnov2003BlackHole} showed that one could use the solution of the Liouville equation \cref{eq:Liouville} to design a one parameter family of regularising surfaces near the boundary such that the regularising surface approaches the boundary as the regulator is removed, with the required uniform negatively curved metric on the boundary surface. 
Using this $\Phi$-dependent regularisation, the gravitational action can be written in terms of the on-shell Takhtajan-Zograf action \cite{Zograf1988Uniformosation} which is a functional of the Liouville field.  
Thus, the gravitational action calculation has been reduced to solving the Liouville field equation as we have described above. The package
\textit{Handlebodies.m}\cite{Wien2017Numerical} makes use of its numerical solution for the Liouville field in order to numerically compute this action.

\begin{figure}[ht]
    \centering
    \begin{subfigure}[b]{0.47\textwidth}
        \centering
        \includegraphics[width=\textwidth]{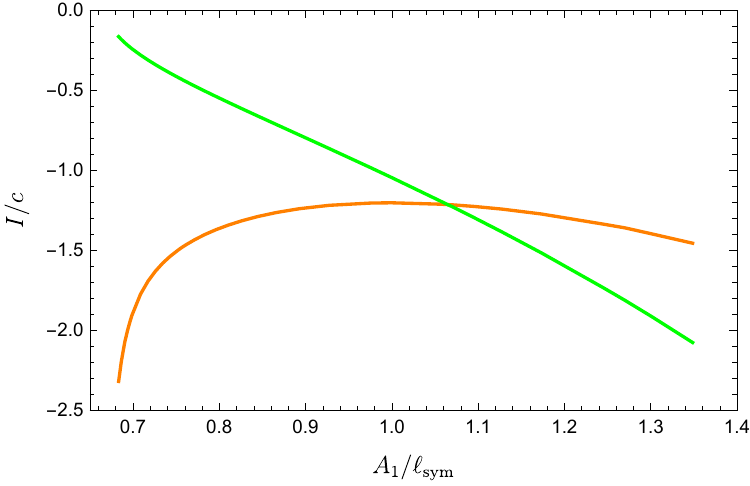}
        % \caption{Subfigure 2}
        \label{fig:ActA2}
    \end{subfigure}
    \hfill
     \begin{subfigure}[b]{0.47\textwidth}
        \centering
        \includegraphics[width=\textwidth]{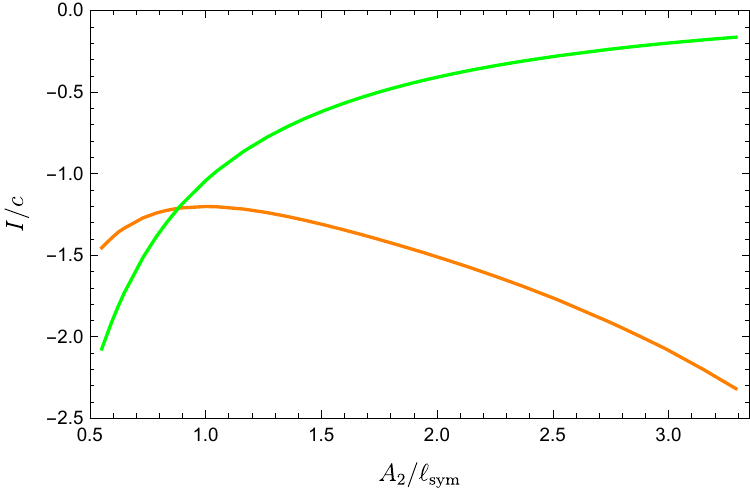}
        % \caption{Subfigure 1}
        \label{fig:ActA1}
    \end{subfigure}
    \caption{Gravitational action of the two possible genus 2 saddles is plotted as a function of the length of the cycles fixed by reflection symmetry. 
    In both figures, the cosmological action is shown in green and the non-cosmological action is plotted in orange. The left and right panel show the actions respectively as a function of the cycle lengths $A_1$ and $A_2$, in units of $\ell_{\mathrm sym}$, the cycle lengths  of the maximally symmetric genus 2 manifold, the Bolza curve. }
    \label{fig:ActionCompare}
\end{figure}
As shown in \cref{fig:ActionCompare},  the action of the cosmological saddle decreases monotonically with the length of the $A_1$ cycle (i.e., increases with the $A_2$ cycle). On the other hand the non-cosmological saddle action has a local maximum. 
 There is a point where both the actions are equal; the value of the action here is $-1.2133c$, where $c = 3 \ell_{AdS}/2G$ is the central charge of the CFT. An interesting point is that this Riemann surface does not correspond to the maximally symmetric Bolza surface, but has moduli close to it.
 
Our results agree with those reported in \cite{Maxfield2016HigherGenus}. In their paper they explore the full space parameterised by $\alpha$ and $\lambda$ and create a phase diagram (their Fig. 7) showing the subspaces where the various saddles dominate. In the language of that paper, our cosmological saddle corresponds to the ``partially connected'' phase, while the ``connected'' and ``disconnected'' phases correspond to our non-cosmological saddles. We are only exploring the 1 parameter family where the $A$ and $B$ cycles are of equal length. This follows the curve between the connected and disconnected phases in Fig. 7 of \cite{Maxfield2016HigherGenus} and then continues into the partially connected phase. The triple intersection point in their figure corresponds to the first order phase transition between the non-cosmological and the cosmological transition in our case, and we find parameter values for this transition in agreement with those in that figure.

For the case $n=1$ (corresponding in the cosmological case to a toroidal universe with a single black hole), the genus two action difference is the only contribution to equation (\cref{eq:nncomp}), so we can say that the cosmological saddle dominates for $\rho < 0.087 $ or for black hole horizon length $R > 22.341\ell$.

\begin{figure}
    \centering
    \includegraphics[width=0.98\linewidth]{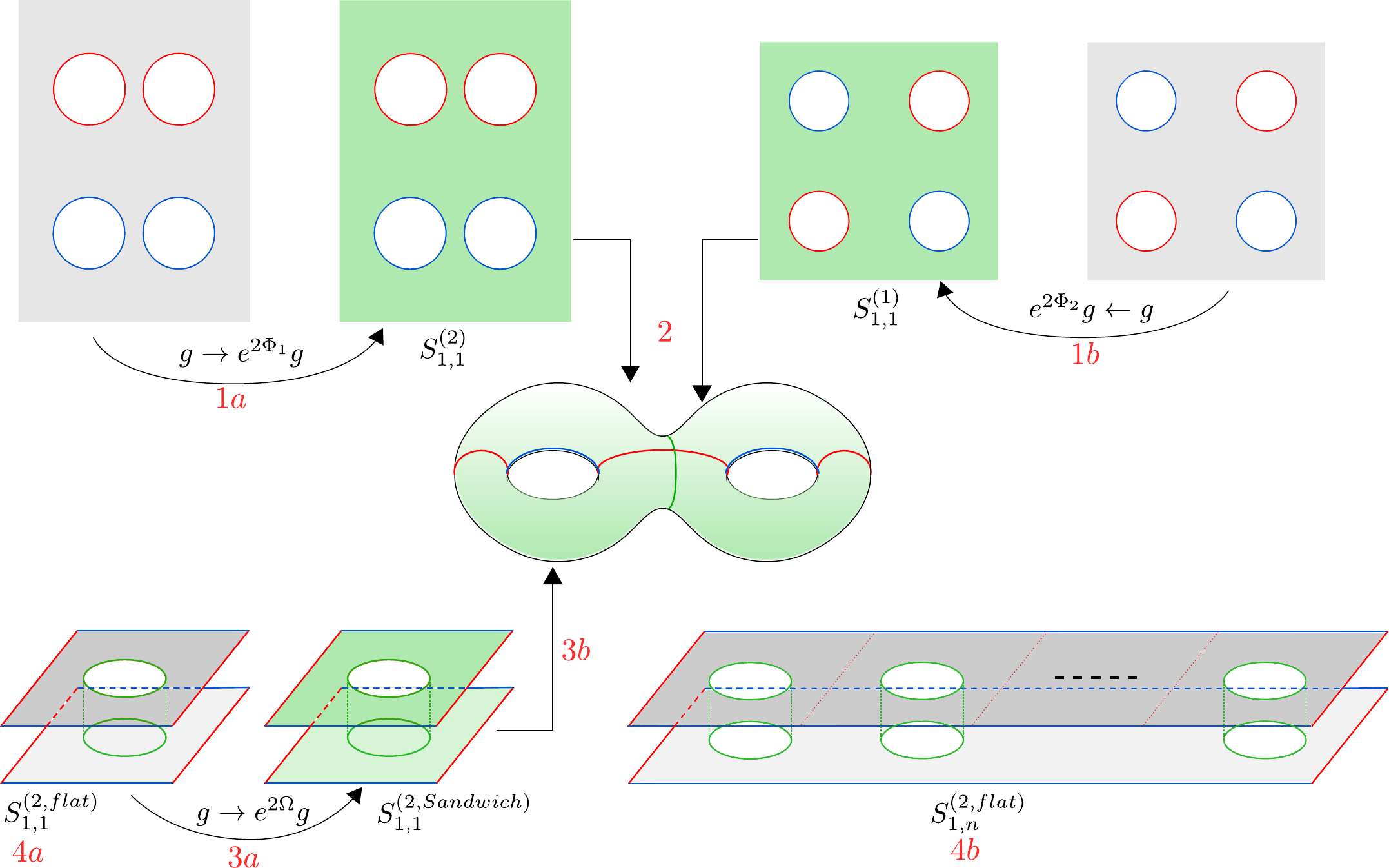}
    \caption{ A step-by-step guide to the numerical computations. The green coloured manifolds have a constant negative curvature whereas the grey ones have a flat metric. 
    First we calculate the Weyl factors $\Phi_1, \Phi_2$ in the Schottky representation of the boundary to obtain a uniform negatively curved metric. In
    step 2 we match the cycle lengths of the genus 2 manifold thus obtained to find the Schottky parameters $\alpha,\lambda$ as a function of $\rho$, and compare the actions $S^{(1)}_{1,1},S^{(2)}_{1,1} $. 
    This gives us the first difference in \cref{eq:nncomp}. 
    In step 3, we find the canonical metric on the sandwich manifold and calculate the cycle lengths to find the parameter $\cal R$ as a function of $\rho$ that correspond to the same boundary manifolds obtained in Step 2. 
    Finally in step 4, we calculate the action $S^{(2)}_{1,1}$ in the flat conformal frame by calculating the Weyl anomaly and subsequently calculate the second term in \cref{eq:nncomp}.
    }
    \label{fig:flowchart}
\end{figure}

The numerical procedure that we have described above allows us to compute the  first difference in brackets in \cref{eq:nncomp}. 
We shall use a similar approach for calculating the second term in \cref{eq:nncomp} as well, although in a different conformal frame. 
For the convenience of the reader we have outlined the organisation of our numerical calculations in \cref{fig:flowchart}, showing different conformal frames and boundary geometries involved.

\subsection{Replica contribution}
\label{subsec:Replica}

In this section, we consider the remaining contribution to \cref{eq:nncomp}, relevant for $n>1$. We would like to understand the behavior of 
\begin{equation}
\label{repdif}
    (S^{(2)}_{1,1} - {1 \over n} S^{(2)}_{1,n})
\end{equation}
as a function of $\rho$ (or $R$) and $n$.

\begin{figure}
    \centering
    \includegraphics[width=0.7\linewidth]{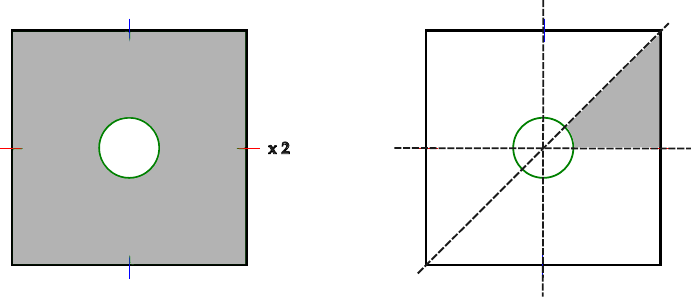}
    \caption{The boundary geometry used to calculate the replica contribution is shown on the left. It consists of a pair of square tori (with opposite sides identified) glued along a circular hole 
    shown in green) at the center of the square. Each toroidal half has a $\mathbb{Z}_2\times\mathbb{Z}_2\times\mathbb{Z}_2$ symmetry as shown in the right. Combined with the exchange symmetry exchanging the top and bottom half, the geometry has a $\mathbb{D}_4$ symmetry. }
    \label{fig:sandwich}
\end{figure}

In a particular conformal frame, the boundary geometry for $S^{(2)}_{1,1}$ can be taken to be the one shown in \cref{fig:sandwich}, with a pair of flat square tori glued along the boundary of a circle removed from each of the tori.\footnote{There will be a cusp at the location of the circle with this gluing. In \cref{app:Schottky} we outline an approach for obtaining the sandwich manifold from the Schottky descriptions considered in \cref{subsec:Genus2term}. 
} 
Let us call this the sandwich manifold. 
This boundary geometry preserves the $\mathbb{D}_4$ symmetry, so should correspond to the cosmological saddle with some value of $\rho$. 

In general, the non-cosmological saddle geometry associated with $S^{(2)}_{1,n}$ is difficult to determine, but simplifications occur either when the gluing circles are very small or when we take large $n$. In this case, we can make use of the following approximation. 
We consider first the geometries obtained by ignoring the circles. In this case, we have a pair of rectangular tori with side lengths $L$ and $nL$. Each of these has a bulk saddle where the circle of length $nL$ is contractible. The geometry is reviewed in Appendix \ref{app:torus}. 

Now consider again the $n$ gluing circles on the boundary of each of these filled tori, and consider the extremal surfaces in the bulk associated with each of the gluing circles. For small circles or large $n$, these extremal surfaces will lie in a region that is well approximated by Poincar\'e-AdS. Extremal surfaces with a circular boundary in Poincar\'e-AdS have zero extrinsic curvature. Thus, if we remove the half-ball regions outside the extremal surfaces for each of the gluing circles, we can glue the two filled tori together along these extremal surfaces; the gluing is smooth in the limit where the extremal surfaces have vanishing extrinsic curvature. For finite $n$ and finite circle size, the gluing won't be quite smooth, and the geometry won't quite be a solution to Einstein's equations, but we can take the action for this approximate solution as an approximation to the correct action. 

For this approximate saddle, we can write the action as
\begin{equation}
\label{eq:TubeApprox}
    S^{2,approx}_{1,n} = 2 S^{torus} - 2n S^{half \; ball} \; .
\end{equation}
As we show in \cref{app:torus}, the  regularized action for the torus is given by
\begin{equation}
\label{Storus}
S^{torus} = - {\ell \pi \over 4 G}{1 \over n}
\end{equation}
while the regularized action for each half-ball with boundary circle radius ${\cal R}$
\begin{equation}
\label{Shalfball}
S^{half \; ball} = - {\ell \pi \over 4 G} \ln {{\cal R} \over \epsilon} \; .
\end{equation}
Combining these, we have that
\begin{equation}
    S^{2,approx}_{1,n} = {\ell \over 2 G} (-{\pi \over n} + n \ln {{\cal R} \over \epsilon})
\end{equation}
The cutoff dependence here will cancel with a similar dependence in the first term of (\ref{repdif}). All together, we have
\begin{equation}
\label{eq:TubeContribution}
    (S^{(2)}_{1,1} - {1 \over n} S^{(2)}_{1,n}) \approx \left[ S^{(2)}_{1,1} - {\ell \over 2 G} \ln {{\cal R} \over \epsilon} \right] + {\ell \over 2 G} {\pi \over n^2}
\end{equation}
Here, the $n$ dependence is all in the second term and is such that the expression decreases with $n$. Since the difference on the left is exactly 0 for $n=1$, we expect it to be negative and decreasing with $n$, at least in the situations (large $n$ for general ${\cal R}$ or small ${\cal R}$ for general $n$) where our approximation is reliable. 

For small ${\cal R}$, the approximation (\cref{eq:TubeApprox}) should be good for $n=1$, so we can write the more explicit expression
\begin{equation}
    (S^{(2)}_{1,1} - {1 \over n} S^{(2)}_{1,n}) \approx  {\pi \ell \over 2 G} \left(-1 + {1 \over n^2} \right)
\end{equation}
This is independent of ${\cal R}$ and should give the precise limit of the more general expression as ${\cal R} \to 0$.

For a general $\cal R$ and $n=1$, the approximation \cref{eq:TubeApprox} is not valid. So we use a different approach to calculate $S^{(2)}_{1,1}$ which does not require us to know how the top and bottom half of the bulk spacetime are joined in the $n=1$ case. 
Our strategy is the following. First, we find the required Weyl transform of the flat metric to impose a constant negative curvature metric on the sandwich manifold.
This allows us to match the boundary geometry to a corresponding Schottky description discussed in \cref{subsec:Genus2term}. 
The action corresponding to these two manifolds should also be equal upto a constant. 
We then relate the action calculated in the Schottky description to the action in the flat conformal frame required in \cref{eq:TubeContribution} by adding a Liouville term evaluated on the Weyl factor calculated above. 
We describe this procedure in detail below.

We obtain the Weyl transformation $\Omega$ by solving the Liouville differential equation 
\begin{equation}
\label{eq:Liouville2}
    \nabla_a\nabla^{a} \Omega = e^{2\Omega}
\end{equation}
over the sandwich manifold shown in \cref{fig:sandwich}. Due to the exchange symmetry between the top and bottom half of this picture, it is sufficient to solve this equation on one half, which we take to be a square bounded by the lines $x=\pm1$ and $y=\pm1$ (in AdS length units) with a circle of size $\mathcal{R} \in (0,1) $ removed from the centre. 
We further reduce the domain to one-eighth of its initial size using the three reflection symmetries; one each along a diagonal and the coordinate axes. 
The resulting domain is a quadrilateral bounded by three straight lines (say, for concreteness, the $x-$axis, the line $x=1$ and the diagonal $x=y$) and a circular arc of length $\pi/4$ (shown on the right hand side panel of \cref{fig:sandwich}).

In the locally flat conformal frame the manifold has a cusp at the circular hole. We see this cusp clearly in quantities such as the circumference of the circles of radius $r$, in this case is simply $2\pi r$, which is non-differentiable at $r=\mathcal{R}$, as its radial derivative changes discontinuously from $+2\pi$ to $-2\pi$ while crossing the hole.
The boundary conditions satisfied by $\Omega$ are chosen such that the resulting Weyl transformed manifold is smooth everywhere. 
Smoothness at the circular hole requires the proper length of any arc of size $\delta$ to vary smoothly at the circular hole. 
This gives us the condition\footnote{These boundary conditions yield the area given by the Gauss-Bonnet formula for 2D Riemann surfaces. Integrating the Liouville equation \cref{eq:Liouville2} over the entire domain, the R.H.S. gives the area of the genus-2 surface. The L.H.S., a total derivative, can be evaluated using integration by parts with Neumann boundary conditions \cref{eq:NeumCirc}, yielding $4\pi$, matching the area from Gauss-Bonnet.
}
\begin{align}
\label{eq:NeumCirc}
    \partial_{\rho}(e^{2\Omega(\rho,\theta)}\rho\delta)\big|_{\rho=\mathcal{R}}=0 \;\; \implies \;\; \partial_r\Omega(\mathcal{R},\theta) = -1/\mathcal{R}.
\end{align}
At all other parts of the boundary the normal derivative of $\Omega$ vanishes due to reflection symmetry. 
Since any genus two surface can be constructed as a quotient of the hyperbolic plane with uniform negative curvature, the Liouville equation has a unique solution given the above Neumann boundary conditions, which we solve for numerically. 
In terms of this Weyl factor the various cycle lengths are given by:
\begin{align}
\label{eq:SandwichCycles}
    C=8\int_{0}^{\pi/4}\dd \theta \mathcal{R} e^{\Omega(\mathcal{R},\theta)}, \quad
    A_2=B_2 = 4\int_{\mathcal{R}}^{1} \dd x e^{\Omega(x, 0)}, \quad
    A_1=B_1 = 2\int_{0}^{1} \dd y e^{\Omega(1, y)}
\end{align}
Thus we obtain a family of genus 2 Riemann surfaces with constant negative curvature and a $\mathbb{D}_4$ automorphism group, parametrised by $\mathcal{R}$.

In \cref{subsec:Genus2term}  we obtained another one-parameter family non-cosmological handlebodies with a genus 2 boundary having a Schottky description in terms of the Schottky parameter $\rho$. We can find a one-one mapping between the parameters $\rho$ and $\cal R$ by matching the cycle lengths of their corresponding Riemann manifolds. 
Both these families of Riemann surfaces, and their corresponding bulk geometries, represent the same set of handlebody spacetimes related by a diffeomorphism that we have not explicitly determined. 
The question of whether their regularised action is the same is a more subtle question as the spacetimes are  regularised differently. 

The regularising surface $z=\epsilon$ used for the sandwich manifold may be mapped by the diffeomorphism into a surface different from the canonical regularising surface (characterised by some regulator $\tilde{\epsilon}$) used in the Schottky description which depends on the other Weyl factor $\Phi$ \cite{Krasnov2000Holography}.
However, since the regulator surfaces must approach the same boundary surface in the limit $\epsilon,\tilde{\epsilon}\rightarrow 0$, $\tilde{\epsilon}$ must be a multiple of $\epsilon$ in this limit. 
The regularised action of a genus 2 handlebody contains a logarithmic divergence $\frac{\ell}{2G}\log{\epsilon}$. 
This divergence has its origin in the conformal anomaly and is universal, independent of the boundary moduli and the conformal frame. 
Thus the action in the Schottky picture and the sandwich picture may differ at most by a constant $k$.
So we have
\begin{equation}
    S^{2,sandwich}_{1,1} = S^{2,Schottky}_{1,1} - \frac{\ell}{2G}\log{\epsilon} + k ,
\end{equation}
where $S^{(2,Schottky)}_{1,1}$ is the finite part of the action in the Schottky picture, calculated numerically in \cref{subsec:Genus2term}. 

\begin{figure}
    \centering
    \includegraphics[width=0.7\linewidth]{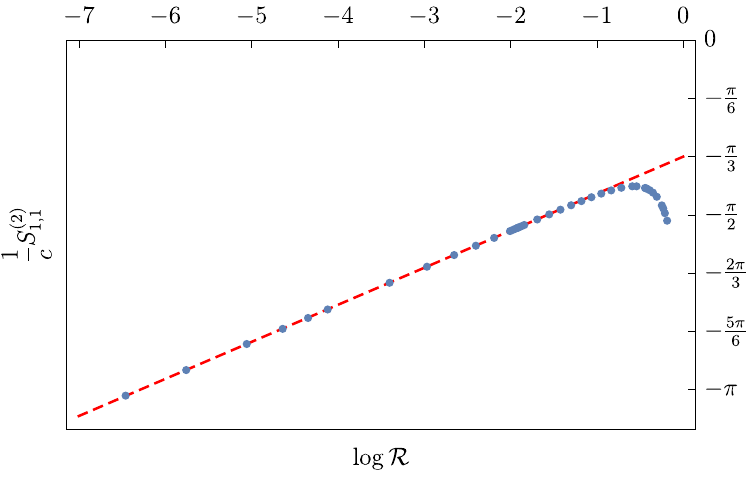}
    \caption{The action $S^{(2)}_{1,1}$ calculated in the flat conformal frame with the sandwich manifold as the boundary, as a function of the log of the hole size $\mathcal{R}$. The blue dots represent numerical values of the action including the constant shift $\alpha=0.063$. The dashed line shows the action calculated analytically using the approximation \cref{eq:TubeApprox}. We see that the approximation holds remarkably well for $\log\mathcal{R}<-1$ or $\mathcal{R}<0.36$.}
    \label{fig:LiouvilleFit}
\end{figure}

The partition function of a 2D CFT under a Weyl transformation of its underlying metric transforms as
\begin{equation}
\label{eq:WeylTransform}
  Z[e^{2\Omega} g] = e^{I_L[g, \Omega]} Z[g] 
\end{equation}
where $I_L$ is the Liouville action 
\begin{equation}
    I_L[g, \Omega] = \frac{c}{24\pi} \int d^2x \, \sqrt{g} \left( \nabla_a \Omega \nabla^a\Omega + \mathscr{R} \Omega \right), 
\end{equation}
and $\mathscr{R}$ is the Ricci curvature associated with the metric $g$.
Applying $\log{}$ to both sides of \cref{eq:WeylTransform}, we see that actions in the flat and the canonical frame are related as
\begin{align}
    S^{(2,flat)}_{1,1} &= S^{(2,sandwich)}_{1,1}+I_{L}[\eta,\Omega] \\
    & = S^{(2,Schottky)}_{1,1}+I_{L}[\eta,\Omega]-\frac{\ell}{2G}\log{\epsilon} + k.
\end{align}
We evaluate the R.H.S of the above equation numerically and the chose the constant $k$ to fit the analytical small $\cal R$ limit,
\begin{equation}
     S^{(2,flat)}_{1,1}(\mathcal{R}) = \frac{\ell}{2G}\left( -\pi +\log{\frac{\mathcal{R}}{2\epsilon}} \right).
\end{equation}
We present our numerical calculation in \cref{fig:LiouvilleFit} and find the constant $k \approx 0.063.$

\begin{figure}[ht]
    \centering
    \includegraphics[width=0.6\linewidth]{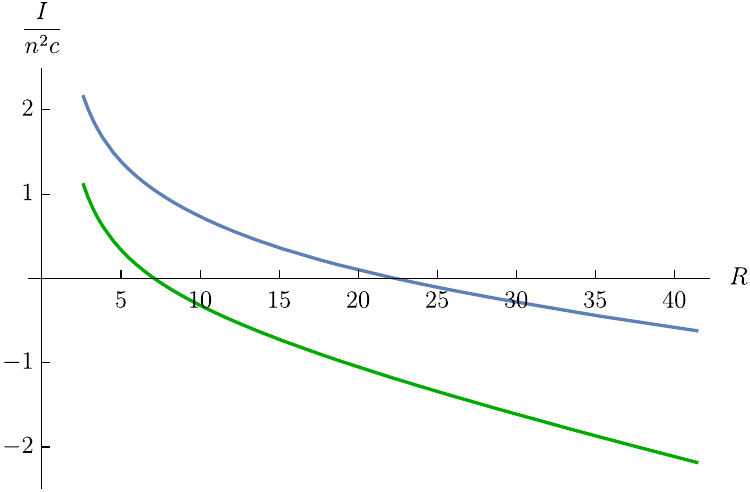}
    \caption{Action difference per lattice site between the cosmological saddle and the non-cosmological saddle, plotted against the event horizon length of the black holes in the cosmological saddle (in AdS units).
    The blue curve shows the action difference for the $n=1$ case. The green curve shows the total action difference per lattice site in the limit $n \to \infty$.}
    \label{fig:FullAction}
\end{figure}

\subsection{Total action difference}

We now consider the complete action difference in (\cref{eq:nncomp}).

As we mentioned above, for $n=1$, only the first term contributes and the cosmological saddle dominates for $\rho < 0.087 $ or for black hole horizon length $R > 22.341\ell $. 
The replica term in \cref{eq:nncomp} is non-zero for $n>1$. From \cref{fig:LiouvilleFit} we see that 
\begin{equation*}
    S^{(2)}_{1,1} < \frac{\ell}{2G}(-\pi + \log{\frac{\cal R}{\epsilon}}).
\end{equation*}
In conjunction with \cref{eq:TubeContribution} this implies that the replica contribution is always less than or equal to $-\pi\ell/2G(1-1/n^2)$, with the equality holding for small $\mathcal{R}$.
Thus the replica contribution is always negative and decreasing with $n$. 
If we increase the size of the lattice (so that we have more black holes) with the other parameters fixed, the cosmological saddle is more likely to dominate. 
Numerically, we find that in the limit of an infinite lattice, the cosmological saddle has lower action when the black hole has a radius above the critical value $R^{crit}_{\infty}\approx7.06\ell$. This is illustrated in  \cref{fig:FullAction}, which shows the total action difference per lattice site for a $1 \times 1$ lattice (blue) and in the limit of an infinite lattice (green).
The critical size corresponds to a black hole horizon radius $r_{crit} = R_{crit}/(2 \pi) \approx 1.12 \ell_{AdS}$, close to the radius $\ell_{AdS}$ of a 3D black hole just above the Hawking-Page transition. From the general relation (\ref{DvsR}) we find that the critical radius corresponds to a critical separation $d_{crit} \approx 1.2 \ell_{AdS}$.

\section{Discussion}

In this paper, we have described a microscopic construction of a big-bang / big-crunch cosmological spacetime using the tools of holography. We used a Euclidean CFT path integral to define a particular entangled state of a collection of uncoupled CFTs; in some cases, the dual spacetime has an interior region that is a cosmology where the matter is a lattice of black holes.

A significant quantitative result is that the saddle giving rise to the cosmology only dominates the path integral for a range of parameters where the size of the black holes and the separation between the horizons is similar to the cosmological scale, $r > r_{crit} = 1.12 \ell_{AdS}$. In this sense, the spacetimes are quite inhomogeneous at the scale of the lattice, but the lattice itself is perfectly homogeneous and isotropic.

The results show that a cosmology with black holes that are much smaller than the cosmological scale (as in our own universe) will only arise from the contributions of a subdominant saddle and thus will be a rare part of the wavefunction in our construction. It would be interesting to understand whether the physics of such a cosmology can still be extracted from the dual CFT state in some way or whether there is a projection on the CFT state that projects onto this cosmology on the gravity side. For example, projection of the CFT state to a high-energy band for each CFT (with energies of order $c$) might largely remove the contributions of the non-cosmological spacetimes for which the CFT energies are of order $c^0$.

While our explicit calculations have been in three-dimensional gravity, we expect that the general picture should be qualitatively the same in higher-dimensions. For example, the case of a spherical ``lattice'' with two black holes reduces to the thermofield double construction, and here the physics is similar in higher dimensions, with the Hawking-Page transition between the two-sided black hole (the ``cosmological'' solution). Constructing solutions in higher dimensions would generally require numerics, but similar numerical computations have been done in the past (see e.g. \cite{bentivegna2018black}).
A simple analytic construction of various cosmologies with matter and black holes was given in \cite{ Sahu2024Singular}. Here, the idea is to start with a homogeneous and isotropic universe with matter and radiation; from this universe, we can remove spherical spatial regions and patch on an interior region of a Schwarzschild black hole, satisfying the Israel junction conditions via a shell of dust.

\begin{figure}
    \centering \includegraphics[width=0.8\linewidth]{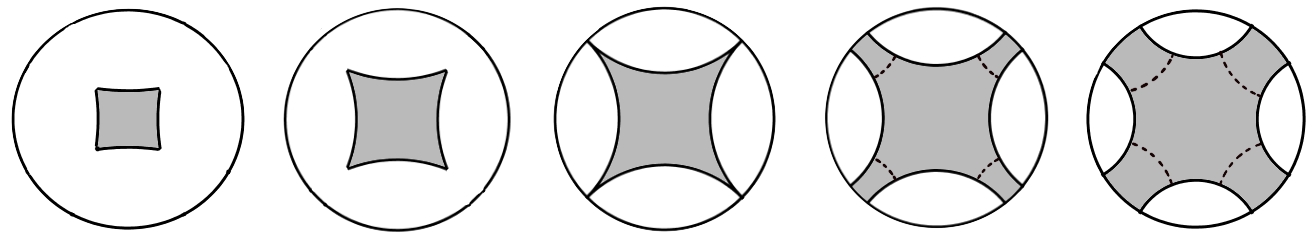}
    \caption{Regions of hyperbolic space giving lattice cells for cosmologies with a square lattice of massive particles of increasing mass (left to right). Below the black hole threshold  (left of center), the corners are associated with conical defects and the solution includes massive particles at these locations. Above the black hole threshold (right of center) the cells include parts of the boundary that give rise to second asymptotic regions associated with the black holes.}
    \label{fig:MassRange}
\end{figure}

\subsubsection*{Insensitivity of scale factor to mass above the black hole threshold}

For the square lattice cosmologies, we have a one- parameter family of solutions parametrized by the particle / black hole mass. All of these solutions have the same scale factor $a(t) = \cos(t/\ell_{AdS})$. The solutions below the black hole threshold (see \cref{fig:MassRange}), described in more detail in \cite{MMV} have a lattice of conical defects each with a particle having mass proportional to the defect angle. The particle mass divided by the area of the lattice cell is the same in all cases, $\rho_M = M/A = 1/(8 \pi G \ell_{AdS}^2)$ and this is precisely the mass density required to produce the scale factor  $a(t) = \cos(t/\ell_{AdS})$ according to the 3D flat space Friedmann equation
\begin{equation}
\label{FriedD}
H^2 + {1 \over \ell_{AdS}^2} = 8 \pi G {\rho_M \over a^2}\;
\end{equation}
for homogeneous and isotropic spacetime. Thus, there are no ``back-reaction'' effects due to the inhomogeneity \cite{MMV}. 
Above the black hole threshold, the area of space excluding the second asymptotic regions is always $2 \pi \ell_{AdS}^2$, independent of the size of the black holes. Thus, if we don't include any space past the black hole horizons when calculating the volume, the mass density keeps increasing while the scale factor evolution is unchanged. If we want the Friedmann equation (\ref{FriedD}) to be satisfied, it is necessary to include a certain additional area associated with each black hole when calculating the density $\rho_M$.

\subsubsection*{A closed universe Hilbert space from auxiliary degrees of freedom}

An interesting feature of our solutions is that while some of them describe ``closed universes,'' e.g. with spherical or toroidal spatial slices, there is still a non-trivial underlying Hilbert space, associated with the second asymptotic regions of the black holes. From the bulk perspective, the cosmology is in a mixed state, purified by fields in the second asymptotic regions. The full states are dual to states in the Hilbert space of the collection of CFTs; here, there are many possible states. 
Thus, our situation appears to be qualitatively different from other settings e.g. \cite{Usatyuk:2024mzs} where it has been argued that the Hilbert space for closed universes is one-dimensional. While our situation has second asymptotic regions for the black holes and thus is in some sense not completely closed, we might expect something similar in cases without black holes where the matter is in a mixed state, entangled with some external auxiliary degrees of freedom.  

\subsection*{Alternative purifications}

In our construction, we began by considering a mixed state cosmology generated by a Euclidean path integral for a pair of Euclidean CFTs with an ensemble of operator insertions. We then purified the construction by noting that the ensemble of insertions could be reproduced by introducing a set of auxiliary degrees of freedom (the CFT degrees of freedom on a set of tubes connecting the two CFTs) and integrating these out. This construction corresponded to a natural purification of the cosmology in the Lorentzian picture where we add a second asymptotic region to each black hole. 

There are many other possible purifications. In the Euclidean picture, this is related to the fact that there are many possible ways to introduce auxiliary degrees of freedom and integrate them out in order to obtain the same ensemble of operator insertions. In various examples, the Lorentzian degrees of freedom associated with the time-symmetric slice of the Euclidean path integral will be different. Thus, the same cosmology can be holographically encoded in many different ways, using many possible sets of degrees of freedom. In the various examples, the spacetime past the horizons of the black holes will be different, but physics in the cosmological region will be the same. The idea that the same spacetime can be encoded using various holographic descriptions was discussed previously in \cite{van2020spacetime,VanRaamsdonk2018,simidzija_holo-ween_2020}.

\subsubsection*{Entanglement wedges and islands}

In cases where the cosmological saddle is dominant, it is interesting to understand how various regions of the cosmology are encoded in the dual CFTs. Consider the case where we have a planar lattice of black holes. Consider the density matrix associated with a collection of CFTs associated with a connected region containing $n$ black holes. A candidate RT surface for the entanglement entropy of this collection of CFTs is the set of $n$ horizons homologous to the $n$ asymptotic boundaries corresponding to the CFTs. This surface has length $n L$. 
However, we can consider an alternative surface in the cosmological region of the geometry that surrounds all of these horizons. The length of this can be upper bounded by $mA$, where $A$ is the distance shown in figure 5 (generally also of order $L$). This will certainly be smaller than the other candidate surface when
\begin{equation}
    mA < nL \; .
\end{equation}
Since $m$ is the number of lattice points on the perimeter of the chosen region on the lattice and $n$ is the number of lattice points in the interior, this will be satisfied for any sufficiently large region of a fixed shape. In this case, the entanglement wedge of the collection of CFTs will include a portion of the cosmology. We can think of such a region  as a cosmological island \cite{Hartman:2020khs}.

\section*{Acknowledgements}

We would like to thank Jim Bryan, Alex Maloney, Viraj Meruliya, and  Petar Simidzija  for helpful discussions.
We acknowledge support from the National Science and Engineering Research Council of Canada
(NSERC) and the Simons foundation via a Simons Investigator Award.

\appendix

\section{Gravitational action in special geometries}
\label{app:torus}

In this appendix we review the calculation of bulk action of some special geometries in Fefferman-Graham coordinates. 

\subsection*{Locally $AdS_3$ with a flat torus boundary}

Consider a Euclidean 2D holographic CFT on a rectangular torus with sides $L_1$ and $L_2$. We can describe this as periodically identified flat space with coordinates $x_1 \sim x_1 + L_1$ and $x_2 \sim x_2 + L_2$. We have two possible dominant saddles for the dual gravity solution. Each can be written as periodically identified AdS. In Fefferman-Graham coordinates, we can represent these as
\begin{equation}
\label{eq:BTZ1}
   ds^2 = { \ell^2 \over z^2} \left(dz^2 + \left(1 - {z^2 \over z_1^2} \right)^2 dx_1^2 + \left(1 + {z^2 \over z_1^2}\right)^2 dx_2^2 \right)
\end{equation}
and
\begin{equation}
\label{eq:BTZ2}
   ds^2 = { \ell^2 \over z^2} \left(dz^2 + \left(1 
   +{z^2 \over z_1^2} \right)^2 dx_1^2 + \left(1 - {z^2 \over z_1^2}\right)^2 dx_2^2 \right)
\end{equation}
where $z_i = L_i/\pi$. In the first solution, the $x_1$ circle is contractible in the bulk, while in the second solution, the $x_2$ circle is contractible in the bulk. The values of $z_i$ are chosen to avoid a conical singularity at $z = z_i$. 

To compare the action for these two solutions, the simplest approach is to choose a cutoff surface at the same $z_c = \epsilon$ in both solutions (it is important that we are using Fefferman-Graham coordinates here, subtract the regularized actions, and take the cutoff to infinity. For this, it suffices to use the bulk action
\begin{equation}
    S_{bulk} = - {1 \over 16 \pi G} \int d^3 x \sqrt{g} (R - 2 \Lambda) \; ,
\end{equation}
where $\Lambda = -1/\ell^2$.
The bulk equation of motion gives
\begin{equation}
  R_{\mu \nu} - {1 \over 2}  g_{\mu \nu} R - {1 \over \ell^2} g_{\mu \nu} = 0 \; . 
\end{equation}
Contracting indices gives
\begin{equation}
   R  = -{6 \over \ell^2} \; . 
\end{equation}
so we have
\begin{equation}
    S_{bulk} =  {1 \over 4 \pi G \ell^2} \int d^3 x \sqrt{g} \; .
\end{equation}

The action for the $i$th solution, integrated from $z = z_i$ to the cutoff $z = \epsilon$ then gives
\begin{eqnarray*}
    S &=&  {\ell \over 4 \pi G} \int_\epsilon^{z_i} d x_1 dx_2 dz ({1 \over z^3} - {z \over z_i^4}) \cr
    &=& \left. {\ell \over 4 \pi G} L_1 L_2 (-{1 \over 2 z^2} - {z^2 \over 2 z_i^4}) \right|_\epsilon^{z_i} \cr
    &=& {\ell \over 4 \pi G} L_1 L_2 \left[({1 \over 2 \epsilon^2} + {\epsilon^2 \over 2 z_i^4}) - {1 \over z_i^2} \right]
    \; .
\end{eqnarray*}
We see that the leading divergent term is the same for both solutions. Subtracting the actions and taking the limit $\epsilon \to 0$ gives the finite result
\begin{equation}
\label{ActionDif}
    S_1 - S_2 = {\ell \over 4 \pi G} L_1 L_2 \left[{1 \over z_2^2} - {1 \over z_1^2} \right] = {\pi \ell \over 4 G} \left[{L_1 \over L_2} - {L_2 \over L_1} \right] 
\end{equation}
We see that solution 1 has lower action if $L_1 < L_2$. In other words, the smaller circle prefers to be contractible in the bulk.

We can alternatively make use of the boundary action plus counterterms to render each term finite individually. The Gibbons-Hawking boundary action is 
\begin{equation}
    S_{GH} = - {1 \over 8 \pi G} \int d^2 x \sqrt{h} K \; .
\end{equation}
To calculate $K$, we can write the metric as 
\begin{equation}
    ds^2 = dr^2 + \gamma_{ij}(r) dx_i dx_j 
\end{equation}
where $\rho$ increases towards the boundary and use $K_{ij} = {1 \over 2} {\cal L}_r \gamma_{ij}$ which leads to
\begin{equation}
    K = {d \over d r} \ln \sqrt{\gamma} \; .
\end{equation}
The appropriate coordinate transformation is $r/L = \ln(z_0/z)$. Up to terms that vanish when $\epsilon \to 0$, we have
\begin{equation}
    K = {2 \over \ell}
\end{equation}
Thus, the Gibbons-Hawking action gives
\begin{equation}
    S_{GH} = - {\ell \over 4 \pi G} L_1 L_2  {1 \over \epsilon^2} \; .
\end{equation}
The required counterterm action is 
\begin{equation}
    S_{ct} = {1 \over 8 \pi G} \int d^2 x \sqrt{h} {1 \over \ell} \; \;.
\end{equation}
Up to terms that vanish for $\epsilon \to 0$, this gives
\begin{equation}
    S_{ct} =  {\ell \over 8 \pi G} L_1 L_2  {1 \over \epsilon^2} \; .
\end{equation}
Adding all terms and taking $\epsilon \to 0$, we have
\begin{equation}
    S_i = - {\ell \pi \over 4 G} {L_1 L_2 \over L_i^2}
\end{equation}

Taking the difference in action for the two saddles recovers the result \ref{ActionDif} above.

\subsubsection*{Hemispheres anchored on the boundary}

Consider a Poincar\'e-AdS spacetime and a hemispherical shell of radius $\cal R$ anchored on a circle on the boundary $z=0$.
We can calculate the action for the spacetime enclosed within this shell with the standard counterterms and a cutoff surface that is $z = \epsilon$ in the Poincar\'e geometry. 
The bulk action is
\begin{eqnarray*}
    S_{bulk} &=& {1 \over 4 \pi G \ell^2} \int d^3 \sqrt{g} \cr
&=&  {1 \over 4 \pi G \ell^2} \int_\epsilon^{\cal R} dz {\ell^3 \over z^3} \int_0^{\sqrt{{\cal R}^2 - z^2}} dr 2 \pi r  \cr
&=&{\ell \over 4 G} \int_\epsilon^{\cal R} dz {1 \over z^3} (\mathcal{R}^2 - z^2)  \cr
&=& {\ell \over 4 G} \left({\mathcal{R}^2 \over 2 \epsilon^2} - \ln {\mathcal{R} \over \epsilon} - {1 \over 2}\right)
\end{eqnarray*}
The GH term and counterterm are
\begin{eqnarray*}
    S_{GH} + S_{ct} &=& {1 \over 8 \pi G} \int d^2 x \sqrt{h} \left( - K + {1 \over \ell} \right) \cr
    &=& -{1 \over 8 \pi G \ell} \int_0^{\sqrt{{\cal R}^2 - \epsilon^2}} dr {\ell^2 \over \epsilon^2} 2 \pi \rho  \cr
     &=&  {\ell \over 4 G} \left(-{\mathcal{R}^2 \over 2 \epsilon^2} + {1 \over 2}\right)
\end{eqnarray*}
This gives us the total action
\begin{equation}
    S_{tot} = - {\ell \over 4 G} \ln {\mathcal{R} \over \epsilon}.
\end{equation}

We need to be careful about the locus where the cutoff surface intersects the extremal surfaces. Here, for finite $\epsilon$, the cutoff surface has a cusp, and this gives a singular extrinsic curvature. 

\section{Details of the Schottky construction for genus two surfaces}
\label{app:Schottky}

\begin{figure}
    \centering
    \includegraphics[scale = 0.3]{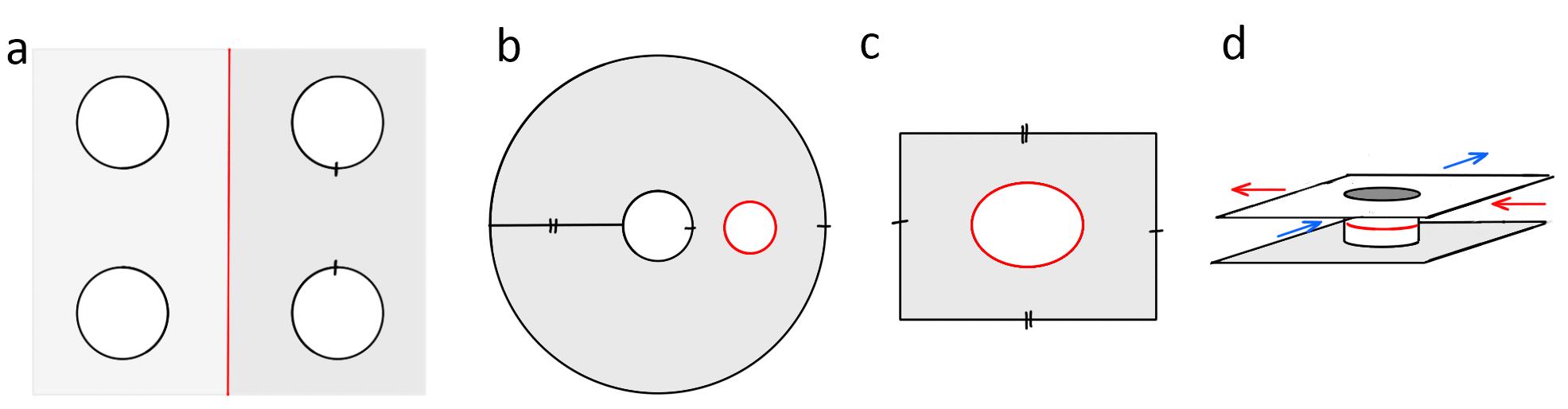}
    \caption{a) Complex plane with pairs of circles identified to define a genus 2 Riemann surface. b) Image of right side after a Mobius transformation. c) Image after a logarithm d) Gluing to another copy (corresponding to the other half of the first figure) gives one ``site'' of the lattice shown in Figure \ref{fig:LatticeBoundary}}  
    \label{fig:Mappings}
\end{figure}

In this appendix, we provide more details of the Schottky construction associated with the non-cosmological saddle. Consider a region of the complex plane obtained by removing unit circles centered at ${u,\bar{u},-u,-\bar{u}}$ where $u = \alpha + i \cosh \lambda$ (Figure \ref{fig:Mappings}a). This region is a fundamental domain for the action of a group with generators $L_1,L_2 \in SL(2, \mathbb{C})$ defined by maps $w \to z$ given by
\begin{equation}
L_1: {z - v \over z - \bar{v}} = e^{- 2 \lambda} {w - v \over w - \bar{v}} \qquad \qquad
L_2: {z + \bar{v} \over z + v} = e^{- 2 \lambda} {w + \bar{v} \over w + v} \qquad \qquad
v \equiv a + i \sinh \lambda 
\end{equation}
Here, $L_1$ maps the exterior of the bottom right circle in Figure \ref{fig:Mappings}a to the interior of the top right circle and $L_2$ maps the exterior of the bottom left circle to the interior of the top left circle. Identifying points by the action of this group, we obtain a genus two surface. We can visualize this by adding a point at infinity so the complex plane topologically becomes a sphere, and then gluing the circles on the right together and the circles on the left together to form two handles. The specific construction we have described gives a family of Riemann genus two Riemann surfaces with $\mathbb{Z}_2 \times \mathbb{Z}_2 \times \mathbb{Z}_2$ symmetry.

One may also show how the construction can be explicitly mapped into a geometry like the sandwich manifold considered in \cref{fig:sandwich}. 
Starting from the right side of \cref{fig:Mappings}a, we can consider a holomorphic transformation 
\begin{equation}
    z \to {v \over |v|}{z-\bar{v} \over z - v} 
\end{equation}
which maps it into the Figure \ref{fig:Mappings}b.
The images of the upper right and lower right circle in \ref{fig:Mappings}a are concentric circles of radius $e^\lambda$ and $e^{-\lambda}$ centered at the origin, while the image of the imaginary axis is a circle of radius $(\sinh{\lambda})/\alpha$ centered at $z= \sqrt{1 + (\sinh^2 \lambda)/\alpha^2}$.

Considering the further map 
\begin{equation}
    z \to \log z 
\end{equation}
with a branch cut taken on the negative real axis maps this to the interior of a rectangle $|\Re(z)| < \lambda$, $|\Im(z)| < \pi$, shown in \cref{fig:Mappings}c.  The imaginary axis in \cref{fig:Mappings}a is now mapped to a curve $C$ described by 
\begin{equation}
\label{Cdef}
    \cosh x = \cos y \sqrt{1 + (\sinh^2 \lambda)/\alpha^2} \; ,
\end{equation}
where $z=x+iy$. 
The lattice site geometry of figure \cref{fig:Mappings}d can be understood as two copies of \cref{fig:Mappings}c glued together along the curve $C$ (thus restoring the left half of \cref{fig:Mappings}a).
The resulting manifold is similar to the sandwich manifold in \cref{fig:sandwich}, except for the fact that the curve $C$ is different from a circle. 
In principle one may consider a mapping, more complicated than the logarithmic mapping, to construct the sandwich manifold starting from the Schottky description.

\section{Automorphisms of genus 2 Riemann surfaces}
\label{app:Automorphism}
Riemann surfaces with $g \geq 2$ admit a discrete group of automorphisms. 
In mathematical literature, these groups have been classified using the properties of the hyperelliptic functions (complex equation of the form $y^2=p(x)$) corresponding to a Riemann surface. 
Hurwitz's theorem provides an upper bound of $84(g-1)$ on the order of the automorphism group for connected Riemann surfaces with $g\geq2$. This upper bound is not always reached however; in case of $g=2$, the biggest automorphism group is of order $48$ and is called a Bolza curve.

We are interested in a two dimensional moduli space of genus 2 Riemann surfaces where the surface has $\mathbb{Z}_2 \times \mathbb{Z}_2 \times \mathbb{Z}_2$ symmetry. 
This 8-fold Abelian group of symmetries is often enhanced at particular points in the moduli space. In this section we want to understand how this enhancement occurs.

A general $g=2$ surface has a moduli space of $3$ complex or $6$ real dimensions. 
When such a surface has a $\mathbb{Z}_2$ symmetry, the number of parameters reduces to $3$, which correspond to the lengths of 3 circles fixed by this $\mathbb{Z}_2$ symmetry. Lets call them \textit{A-cycles}.
Making 2 of these A-cycles equal in length, introduces an additional $\mathbb{Z}_2$ symmetry that swaps the two A-cycles of equal length. 
It is slightly non-trivial to see how, but having two of the $\mathbb{Z}_2$ symmetries ensures a third $\mathbb{Z}_2$ symmetry which fixes a second set of \textit{B-cycles}
\footnote{
The explanation for this is related to the fact that all the genus 2 hypersurfaces under consideration are associated with a hyperelliptic curve of the form $y^2=p(x)$ where $p$ is a polynomial of degree $5$ or $6$. 
Then the involution $y\xrightarrow{}-y$ gives a Riemann surface in which the roles of the A-cycles and the B-cycles are interchanged. Thus we have a different set of moduli, but they correspond to the same genus 2 Riemann surface.
It is similar to the genus 1 case where interchanging the 'contractible' and 'non-contractible' cycles gives a different bulk solution, but the boundary is the same.
Composition of the involution and one of the $\mathbb{Z}_2$ gives rise to a new $\mathbb{Z}_2$.}. Since all these $\mathbb{Z}_2$ symmetries commute the overall symmetry group has a product structure $\mathbb{Z}_2 \times \mathbb{Z}_2 \times \mathbb{Z}_2$. 

If we further consider this 8-fold symmetric surface to have a constant negative curvature (as can be ensured by uniformisation theorems), it can be built from 8 copies of a right angled pentagon as shown in \cref{fig:Riemann2}. 
The two parameters of this surface correspond to two sides of the pentagon. 
Using this pentagon one can show that because two of the A-cycles are equal in length, two of the B-cycles are also equal in length. In fact, the $\mathbb{Z}_2$ symmetry that swaps the two A-cycles also swaps the two equal B-cycles. 

A further reduction in parameter space occurs when all the A-cycles (and consequently the B-cycles) are equal in length. 
Then the 3 A-cycles (or B-cycles) can be permuted, enhancing the $\mathbb{Z}_2$ symmetry associated with exchange of equal A-cycles to a $\mathbb{D}_3$ symmetry; the group of symmetries of a equilateral triangle. 
Hence the automorphism group becomes $\mathbb{D}_3 \times \mathbb{Z}_2 \times \mathbb{Z}_2$. 
It is easy to see that $\mathbb{D}_3 \times \mathbb{Z}_2$ is isomorphic to $\mathbb{D}_6$, the symmetry group of a regular hexagon.
Thus we have a one parameter group of surfaces with conformal automorphism group $\mathbb{D}_6$ and an anti-conformal group $\mathbb{Z}_2$ which can be interpreted as a time-reversal symmetry. 

A different one parameter family arises by introducing a $\mathbb{Z}_2$ symmetry that interchanges the A-cycles and the B-cycles. 
This $\mathbb{Z}_2$ groups acts non-trivially on the original order 8 automorphism group and gives rise to a semi-product structure as we describe below. 
First taking a product of two of the $\mathbb{Z}_2$ groups we get $\mathbb{D}_2$, the dihedral group of order 4, or the group of symmetries of a rectangle. 
$\mathbb{D}_2$ is a normal subgroup of of $\mathbb{D}_4$, the group of symmetry of a square. 
The additional $\mathbb{Z}_2$ that interchanges the A-cycle and B-cycles, can also be thought of as a subgroup of $\mathbb{D}_4$.
This $\mathbb{Z}_2$ group acts on $\mathbb{D}_2$ to give a inner semi-definite product $\mathbb{Z}_2 \ltimes \mathbb{D}_2$ which is nothing but $\mathbb{D}_4$.
Hence the conformal automorphism group becomes $\mathbb{D}_4$ along with a time-reversal anti-conformal automorphism $\mathbb{Z}_2$.

These two one-parameter families meet at a single point, with a 48-fold symmetry
consisting of 24 conformal automorphisms and a time reversal anti-conformal symmetry. The Order 24 group is the Van Dyck group $D(2,4,6).$

\section{Hyperbolic right-angled pentagons}
\label{app:pentagon}

We found that a genus 2 Riemann surface with $\mathbb{Z}_2 \times \mathbb{Z}_2 \times \mathbb{Z}_2$ symmetry and a hyperbolic metric
may be constructed by joining hyperbolic right angled pentagons. 
Such pentagons have only two free parameters; fixing the length of any two adjacent sides fixes the length of the rest of the sides.

Let two adjacent sides of the pentagon be $a,b$ and the unique side non-adjacent to both these sides be of length $c$. Then they satisfy the relation 
\begin{equation}
\label{eq:pentagon_1}
    \cosh c = \sinh a \sinh b,
\end{equation}
which of course holds for any pair of adjacent sides. 
Let the two remaining sides be of lengths $m$, $n$ where the former corresponds to the side between the sides of length $a$ and $c$.
Then the following relation also holds:
\begin{equation}  \sinh{m}\cosh{a}=\cosh{b}\sinh{n}.
\end{equation}
This relation involves 4 adjacent sides and it also holds for any 4 adjacent sides. This set of relations is not independent from the set of relations exemplified by \cref{eq:pentagon_1}; one set of relations can be derived from the other.
Another set of relations involving 3 adjacent sides is
\begin{equation}
    \tanh{m}\cosh{c}\tanh{n} = 1.
\end{equation}

Explicitly, the sides $m,n$ are given in terms of $a,b$ by 
\begin{align}
    \cosh{m} &= \frac{\cosh a \sinh b}{\sqrt{\sinh^2 a \sinh^2 b - 1}} \\
    \cosh{n} &= \frac{\cosh b \sinh a}{\sqrt{\sinh^2 a \sinh^2 b - 1}}.
\end{align}
Thus the three remaining sides of the pentagon are found uniquely in terms of $2$ adjacent sides.

In case two adjacent sides of the pentagon are equal, we have a one parameter family of isosceles pentagons. 
For isosceles pentagons the two sides attached to the equal adjacent sides are also equal. In the notation of the previously described general right angled pentagon, if $a=b$, $m=n$. 
The equal non-adjacent side and the unequal fifth side are given as 
\begin{align}
\label{Hgeom}
    \coth{m} &= \sinh{a} \\
    \cosh{c} &= \sinh^2{a}.
\end{align}
Interestingly there is only one equilateral right pentagon which has a side length of $a = \cosh^{-1}{(1+\sqrt{5})/2}$ in units of the curvature scale.

\bibliographystyle{jhep}
\bibliography{references}

% \section{Useful references}

% \url{https://scientia.mat.utfsm.cl/archivos/vol15/ar8.pdf}

% J. Symbolic Computation (1998) 12, 789–803

% \url{https://www.math.kit.edu/iag3/~herrlich/media/kunming.pdf}

% \url{https://drive.google.com/file/d/1yx_UPcDIsFWw526iqGn_mWWLZ366njL4/view?usp=sharing}

\end{document}